\documentclass[11pt]{article}
\input epsf
\usepackage{epsfig}
\usepackage{fancyhdr}
\usepackage{latexsym}
\usepackage{amsmath}
\usepackage{amssymb}
\usepackage[latin1]{inputenc}
\usepackage{layout}

\newcommand{\gevc}{\ensuremath{{\rm GeV}\!/c}}
\newcommand{\qbar}{\ensuremath{{\rm \bar{q}}}}
\newcommand{\gevcc}{\ensuremath{{\rm GeV}\!/c^2}}
\newcommand{\tevcc}{\ensuremath{{\rm TeV}\!/c^2}}

\newcommand{\sq}{\ensuremath{{\tilde{\rm q}}}}
\newcommand{\glu}{\ensuremath{{\tilde{\rm g}}}}

\newcommand{\msq}{\ensuremath{m_\sq}}
\newcommand{\mglu}{\ensuremath{m_\glu}}

\newcommand{\sqqbar}{\ensuremath{\sq\bar{\rm q}}}
\newcommand{\qqbar}{\ensuremath{{\rm q\bar{q}}}}

\newcommand{\chanc}{\ensuremath{{\rm p p \rightarrow \glu \glu }}}
\newcommand{\chand}{\ensuremath{{\rm p p \rightarrow \sq \glu \rightarrow \glu \glu q }}}
\newcommand{\chane}{\ensuremath{{\rm p p \rightarrow \sq \sq \rightarrow \glu \glu q q}}}
\newcommand{\chancc}{\ensuremath{{\rm g g \rightarrow \glu \glu }}}
\newcommand{\chandd}{\ensuremath{{\rm q \qbar \rightarrow \glu \glu }}}
\def\Journal#1#2#3#4{{#1} {\bf #2} (#3) #4}

\def\NPB{{\em Nucl. Phys.} {\bf  B}}

\def\PLB{{\em Phys. Lett.} {\bf  B}}

\def\PRD{{\em Phys. Rev.} {\bf D}}

\def\EPJ{{\em Eur. Phys. J.} {\bf C}}

\def\MPA{{\em Mod. Phys. Lett.} {\bf A}}
\def\JMA{{\em Int. J. Mod. Phys.} {\bf A}}

\begin{document}
\begin{titlepage}
\mbox{}\vspace{4cm}\mbox{}
\begin{center}
{\Huge\bf
Discovery potential\\ of R--hadrons\\*[4mm]
with the ATLAS detector
}\\
\vspace{1cm}
{\Large A. C. Kraan$^{a}$, J.B. Hansen$^a$, P. Nevski$^b$}\\
\vspace{1cm}
{\it $^a$ University of Copenhagen, Niels Bohr Institute, Blegdamsvej
  17,\\ 2100 Copenhagen, Denmark}\\
{\it $^b$ Brookhaven National Laboratory, PO box 5000, Upton, NY., USA}\\
\end{center}
\vspace{2cm}
\begin{abstract}
The production of exotic heavy hadronic particles arises in several
models for physics beyond the Standard Model.
%JBHA study of the detection possibilities of new heavy hadron states with the ATLAS detector is presented.
The focus is on R-hadrons, which are stable hadronized
gluinos, predicted by certain supersymmetric models. Interactions and signatures of single R-hadrons are studied with the
ATLAS simulation and reconstruction framework.
%JBH and issues concerning triggering on R-hadron events are addressed
The ATLAS fast simulation framework
has been extended to include parameterizations for
R-hadrons. Based on topological and kinematic variables only, the
discovery potential of the ATLAS detector for R-hadron events produced
in \chanc\, is studied for masses below 2~\tevcc. R-hadrons with masses as predicted by standard SUSY scenarios would be discovered already in the very early stages of the running of the LHC. The discovery reach of heavy gluinos, predicted by for example split supersymmetry models, extends up to at least 1.8~\tevcc\ for three years running of the LHC at low luminosity.
\end{abstract}
\end{titlepage}

\pagenumbering{arabic}
\pagestyle{plain}

\section{Introduction}
Heavy stable hadrons are predicted in several extensions of physics beyond the Standard Model. For example, supersymmetry models exist in which the gluino is stable. Models with a stable gluino are reviewed in Ref.~\cite{gunion}, and include gauge mediated supersymmetry breaking models and string motivated supersymmetric models. Recently, stable gluinos with masses of the order of the TeV scale are predicted in the context of split supersymmetry models~\cite{arkani,giudice}. A stable gluino would hadronize into heavy (charged and neutral) bound states. These bound states (for example \glu g, \glu \qqbar, \glu qqq, \sqqbar, \sq qq) are generically called R--hadrons, where the ``R'' refers to the fact that they can only be stable hadrons if R--parity is conserved~\cite{Farrar:1978xj}. Several models exist for the description of the interactions of stable hadrons in matter~\cite{gunion, aafke}. The phenomenology of stable gluinos has been studied previously in Refs.~\cite{gunion,aafke,tilman,hewett,mafi,tobe}. 

In addition to supersymmetry, other extensions of the Standard Model have been proposed, which predict the existence of new heavy stable hadrons, either due to the presence of a new conserved quantum number, or because the decays are suppressed by kinematics or couplings. Examples are theories with leptoquarks~\cite{Leptoquarks}, theories with universal extra dimensions~\cite{Appelquist:2000nn}, theories with new Standard Model fermions~\cite{Frampton:1998up} and certain unification models \cite{Ingelman:1986dp}.

Various searches for stable massive particles have been performed. A comprehensive summary may be found in Refs.~\cite{pdg,Perl:2001xi}. The negative results of cosmic ray searches and searches in matter on earth suggests that these particles would not be strictly stable, so that all cosmological bounds can be evaded. In this note, a particle is referred to as stable as long as it does not decay in the detector. The mass range from accelerator searches is limited to the center of mass energy. In view of the existing mass limits, this study focuses on particles with masses~$\gtrsim$~100~\gevcc. 

%Several models exist for the description of the interactions of stable hadrons in matter~\cite{gunion, aafke}.
The detection of heavy non-hadronically interacting charged particles with the ATLAS detector has earlier been discussed in Refs.~\cite{nisati,giacomop}. In this paper, an overview is presented on the interaction and detection of \emph{hadronically interacting particles} in the ATLAS detector, with the focus on R-hadrons. Following the interaction model as proposed in Ref.~\cite{aafke}, we discuss issues related to triggering, and study R-hadron signatures in the inner detector, calorimeters and muon chambers based on full simulation.
%The results, generically applicable to any kind of heavy particle interacting hadronically, cover the signatures of single R-hadrons in the Transition Radiation Tracker (TRT), calorimeters and muon chambers. Furthermore, some aspects related to triggering will be addressed. 
%Using topological and kinematical variables,
The discovery potential of the ATLAS detector, from an excess of events containing R-hadrons, is investigated for gluinos with masses up to 2~\tevcc, using the standard ATLAS detector reconstruction at low luminosity (10~$\rm fb^{-1}$/year). More details about the simulation and analysis of R-hadrons can be found in Ref.~\cite{aafkethesis}.

The organization of this note is as follows. The production of R-hadrons at the LHC is described in Section~2. Section~3 summarizes the event generation for R-hadron and background events. Possible trigger signals useful for R-hadron selection are discussed in Section~4. Section~5 presents a study of the most important signatures from events with R-hadrons in the ATLAS detector and describes the final event selection. The resulting mass limits are discussed in Section~6. Section~7 addresses specific signatures for R-hadron identification and classification, which may also further extend the discovery reach of the ATLAS detector. Finally a summary is presented in Section~8.

\section{R-hadron production}
Gluino production is a purely strong process and depends, besides the gluino and squark masses, only on the strong coupling constant. A large number of gluinos (roughly $10^8$ for \mglu=100~\gevcc, $10^3$ for \mglu=1~\tevcc) is expected in one year of LHC running at low luminosity (10~$\rm fb^{-1}$/year).
\subsection{Production channels}
\begin{figure}[h]
\begin{picture}(110,130)
\put(13,65){\epsfxsize68mm\epsfbox{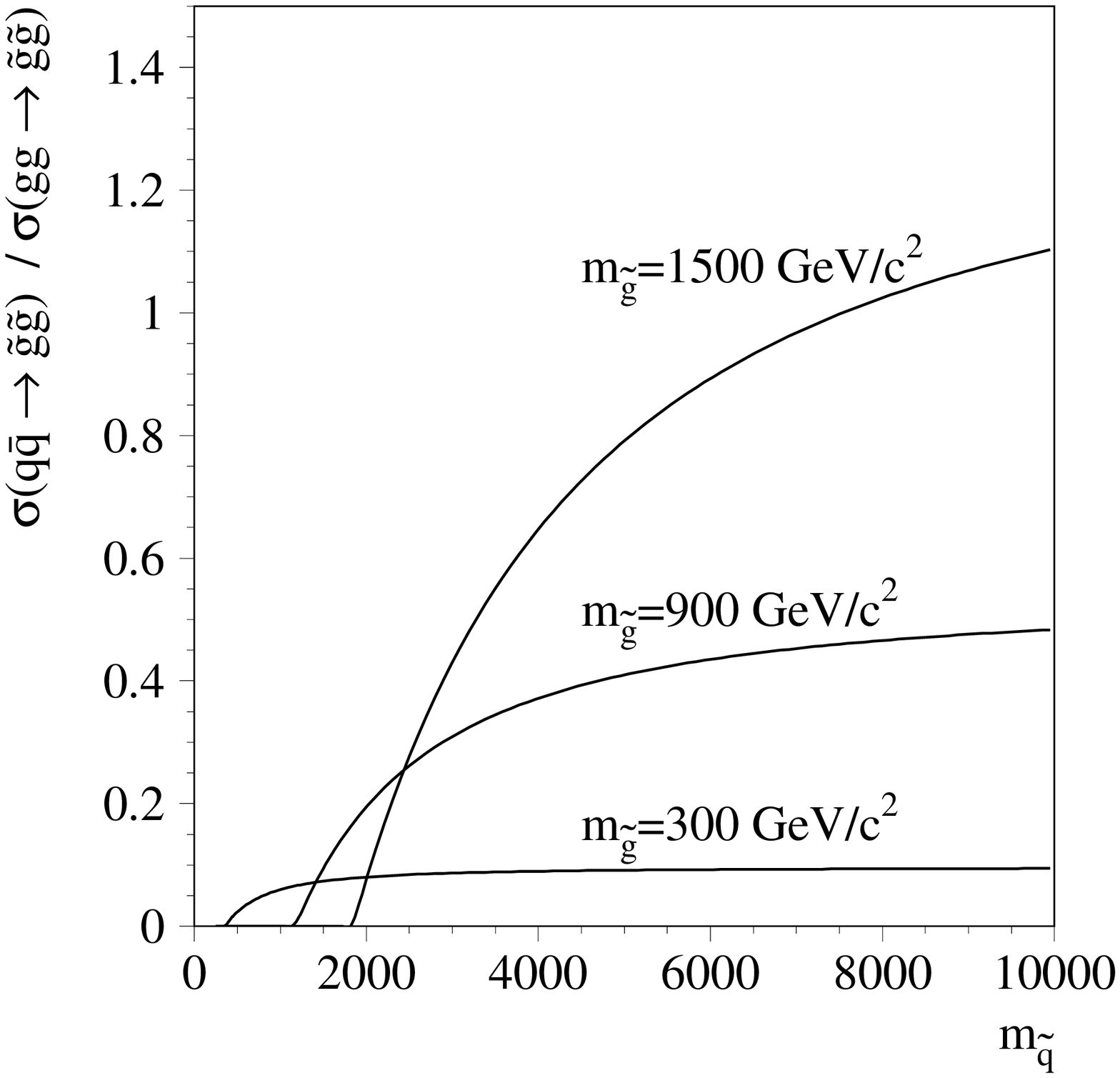}}
\put(30,105){(a)}
\put(80,65){\epsfxsize68mm\epsfbox{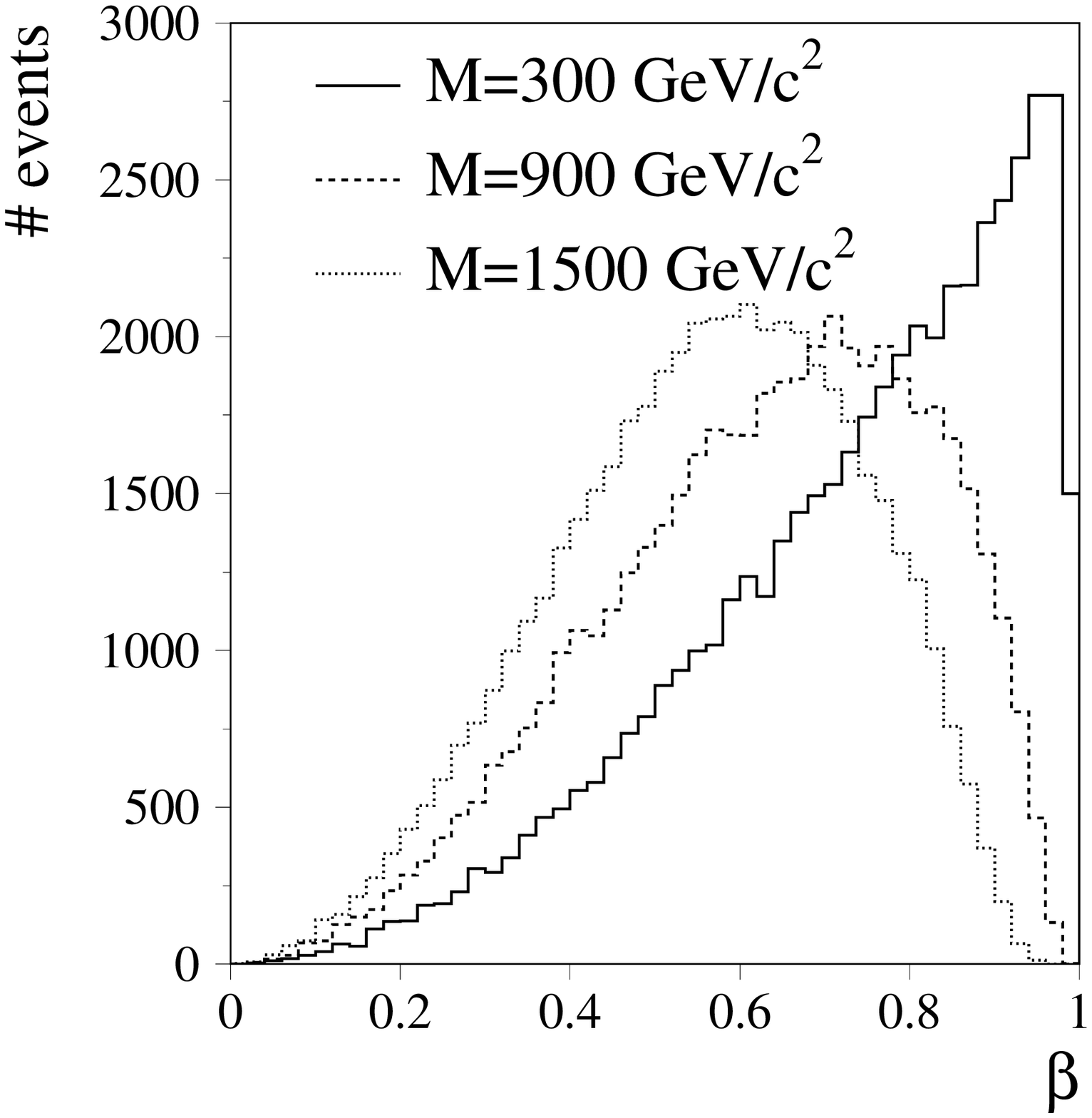}}
\put(100,105){(b)}
\put(13,0){\epsfxsize68mm\epsfbox{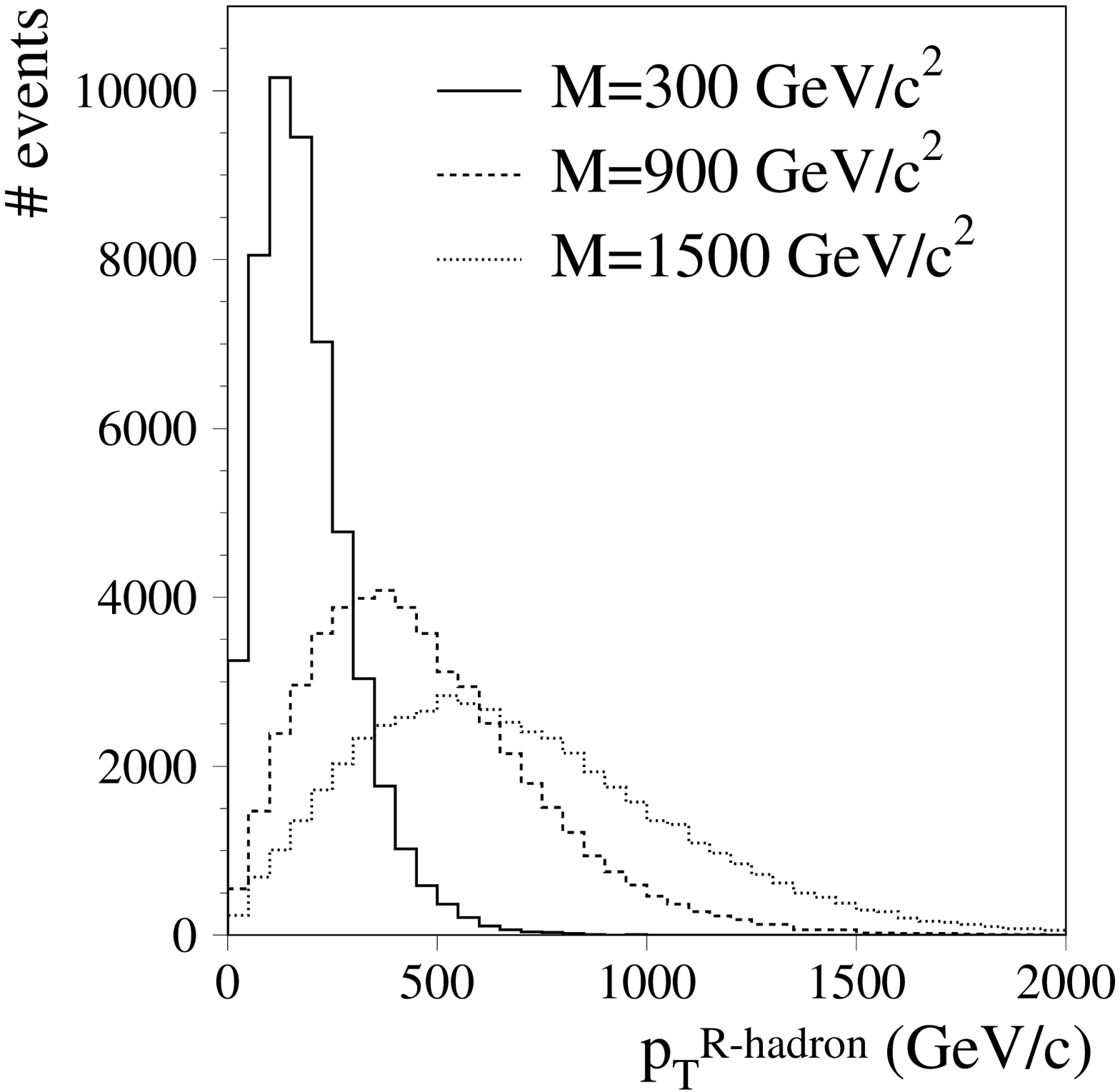}}
\put(60,35){(c)}
\put(80,0){\epsfxsize68mm\epsfbox{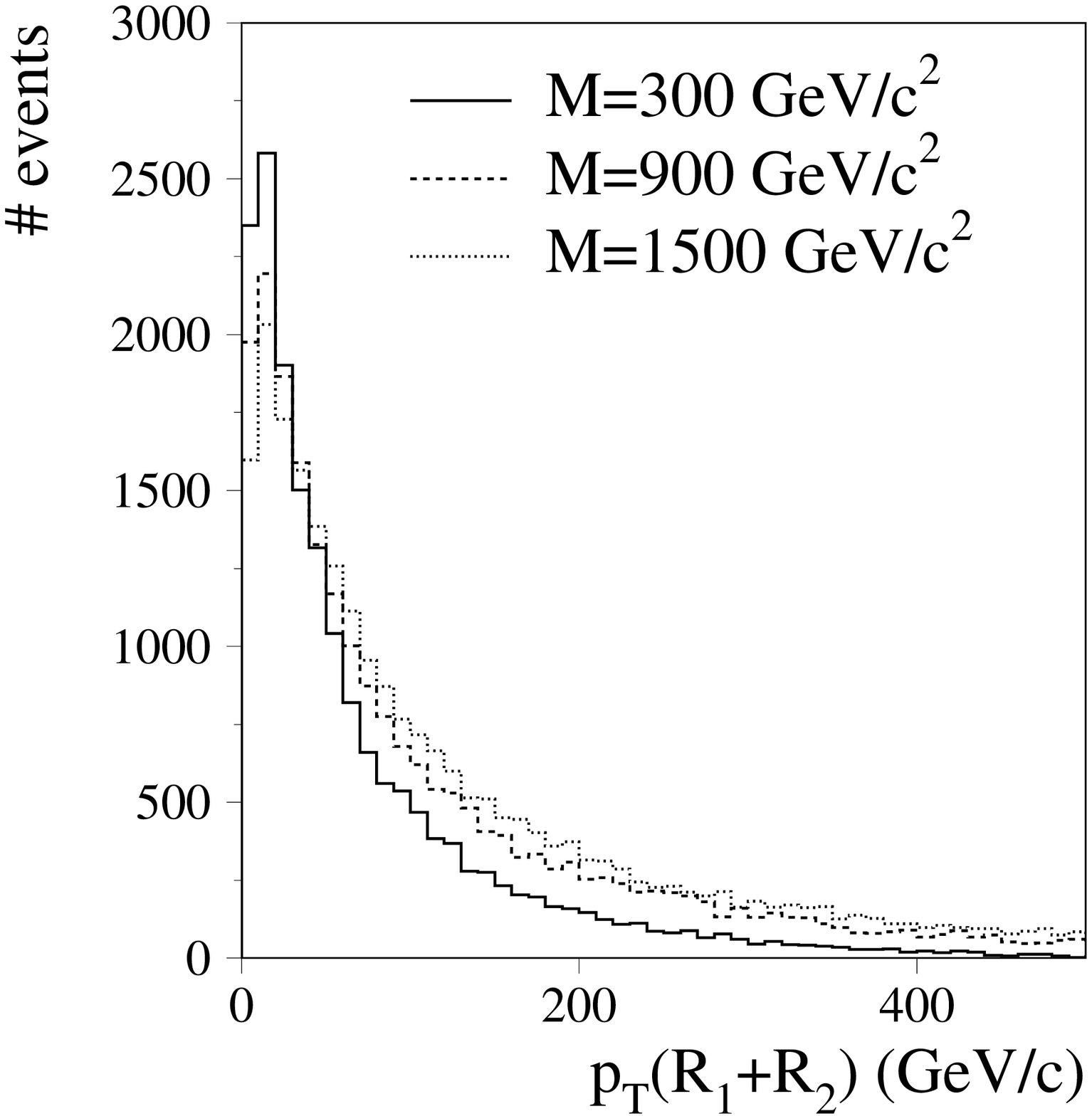}}
\put(128,35){(d)}
\end{picture}
\caption[ ]
{\protect\footnotesize  Aspects of R-hadron production. (a) The ratio of the contribution from \qqbar\ to that of ${\rm gg}$ at the LHC as function of \mglu\ and \msq. (b) The distribution of the velocity of the R-hadrons produced at the LHC for \mglu\ = 300, 900, 1500~\gevcc. Average values are 0.74, 0.64 and 0.56 (c)  The distribution of the transverse momentum of the R-hadrons produced at the LHC for \mglu\ = 300, 900, 1500~\gevcc, with average values 188, 427 and 699~\gevc. (d) The vector sum of the transverse momenta of the two R-hadrons, with average values of 74, 129 and 172~\gevc. For the distributions in (b), (c) and (d) only R-hadrons with $|\eta |<2.5$ were selected.\label{fig:rhadprod}}
\end{figure}

At the LHC, gluinos will be produced via one of the processes \chanc, \chand, or \chane, where the latter two channels are only accessible if the squark is not too heavy. For \glu\glu\ production, the leading channels are $\rm q\bar{\rm q} \rightarrow  \tilde{\rm g}\tilde{\rm g}$ and $\rm g\rm g \rightarrow \tilde{\rm g}\tilde{\rm g}$. In this study, we focus on the channel \chancc, because this process is purely strong and, given the incoming beams and energies, depends only the gluino mass and the strong coupling constant. For moderate \mglu\ (up to $\sim$1400~\gevcc) this channel is dominating when considering \chanc. The ratio of the contribution to $\tilde{\rm g}\tilde{\rm g}$ production from \qqbar\ to that of ${\rm gg}$ at the LHC is shown in Fig.~\ref{fig:rhadprod}a as function of \mglu\ and \msq.
\subsection{Gluino fragmentation}
Long-lived gluinos hadronize into color singlet bound states. In the hadronization process, charged and neutral R-hadrons are produced (\glu q\qbar, \glu qqq, \glu g). For the detection of R-hadrons, 
it is crucial to know which hadron state is the lightest, since that state would dominantly be present in the detector (if it is indeed significantly lighter than the other states). The detection of neutral hadrons differs significantly from that of charged hadrons. The lightest hadron states turn out to be the R-mesons~\cite{aafke,chan,far,fos}, which are shown to be nearly mass degenerate, and slightly lighter than gluino-gluon states. R-baryons are also expected to be degenerate~\cite{aafke,far}, and are roughly 0.3~\gevcc\ heavier than the R-mesons. Thus, there is no preference for an R-hadron to be charged or neutral. The probability to form a \glu g bound state is a free parameter, taken to be 0.1 by default~\cite{torbjorn}. No \glu\glu~-states are considered in the present study and thus R-hadrons are always produced in pairs. For the remaining fraction, the probability to form a neutral and charged R-hadron is about half-half. Only a small amount of baryons ($\sim$2\%) are formed due to baryon suppression. In the present study, gluino hadronization is performed using dedicated PYTHIA routines~\cite{torbjorn}.

\subsection{Event topology}\label{sec:topology}
The combination of phase space, matrix elements and parton density functions results in heavy particles being produced at a $\textrm{p}-\textrm{p}$ collider like the LHC with high $p_T$ values, typically of the order of their own mass, i.e.~they may be relativistic but their mass is still far from negligible. The R-hadrons produced in the channel \chanc\ events are characterized by two R-hadrons produced approximately in a back-to-back configuration in the transverse plane.
%R-hadrons produced in the channels \chand\, and \chane\, are not back-to-back, and results in a large missing energy signature.
The likely presence of a missing energy signal depends among others on the R-hadron mass, because heavier R-hadrons are generally produced with higher momenta than lighter R-hadrons, and thus may result possibly in a larger imbalance.  The distribution of the velocity and transverse momentum of the R-hadrons produced at the LHC via \chancc~ is shown in Fig.~\ref{fig:rhadprod}b and c for different masses of the gluino. Figure~\ref{fig:rhadprod}d shows the total transverse momentum of the two R-hadrons produced via \chancc.

Due to their slow movement, yet high momentum, and their composite colored structure, with one constituent being ultra heavy, R-hadrons posses very distinct features when they traverse ordinary matter. The typical signatures for the detection of single heavy stable R-hadrons are
\begin{itemize}
\item High transverse momentum for charged R-hadrons.
\item Large ionization in the tracking system, in case the R-hadrons are charged and slow.
\item A characteristic pattern of energy deposition in the calorimeters.
\item A large time-of-flight, measurable with the muon chambers.
\end{itemize}

\begin{figure}[]
\begin{center}
\hspace*{-0.48cm}\epsfig{file=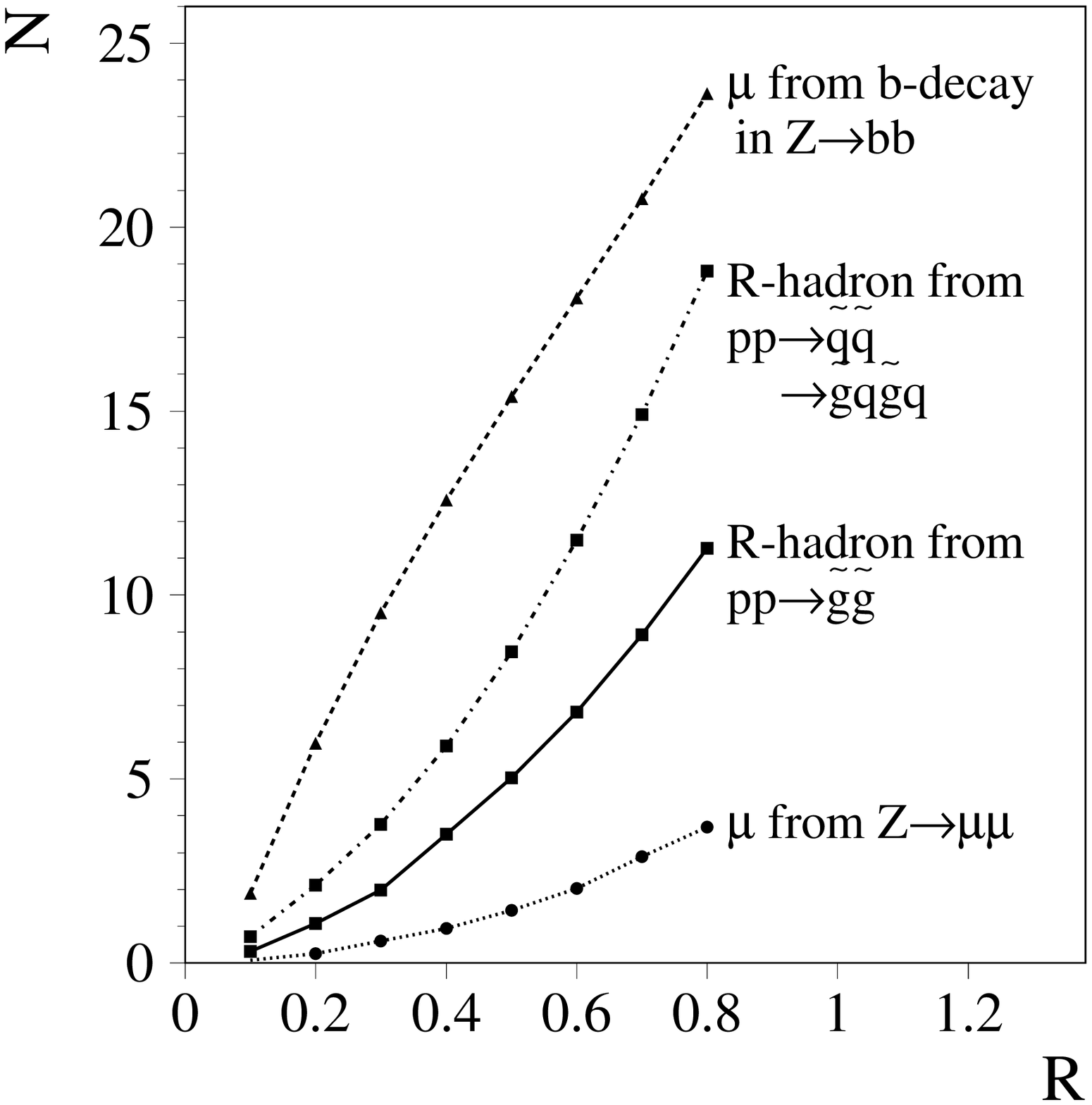,height=5.5cm,width=6.1cm}\hspace*{-0.48cm}\epsfig{file=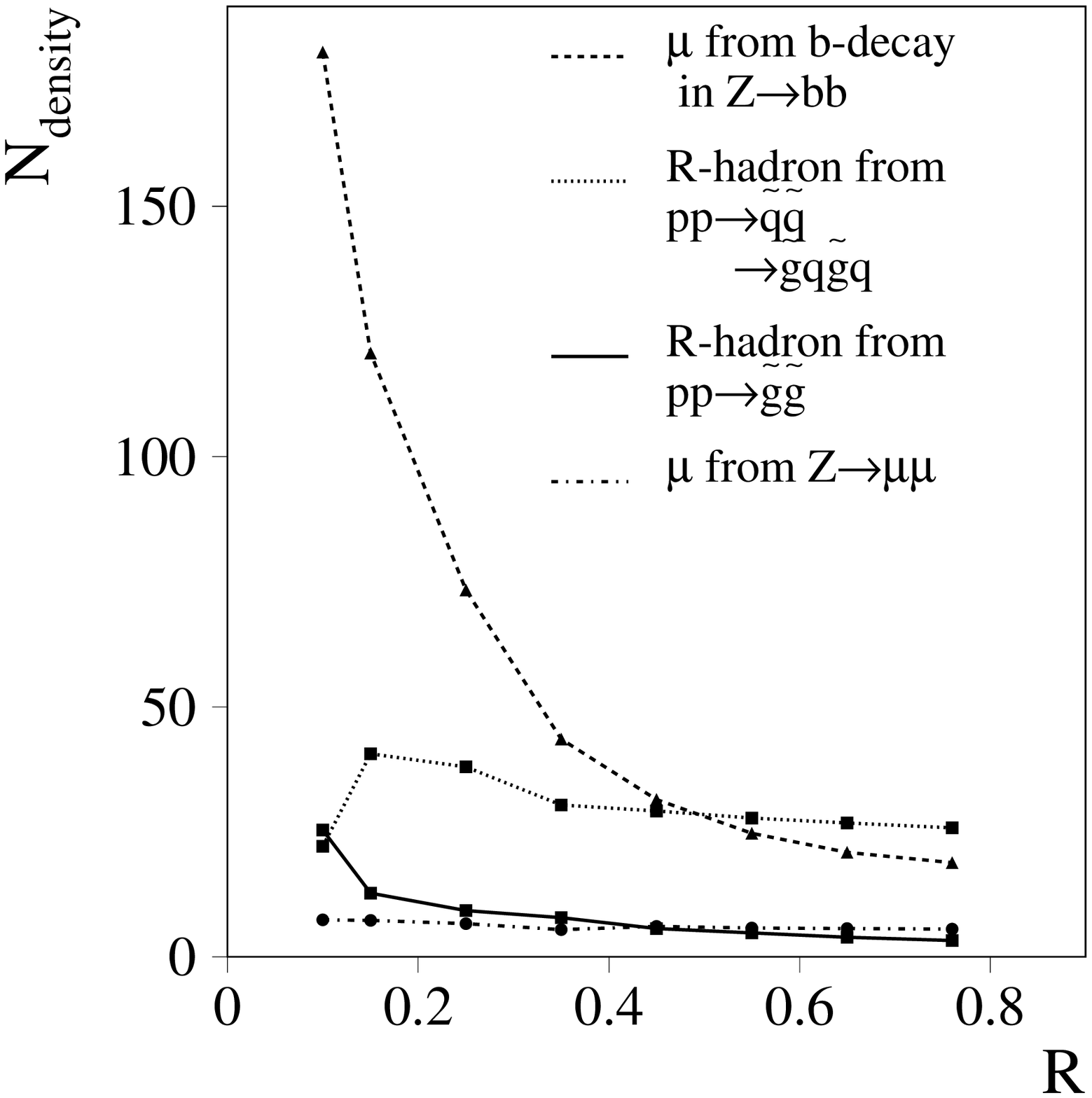,height=5.5cm,width=6.1cm}\hspace*{-0.48cm}\epsfig{file=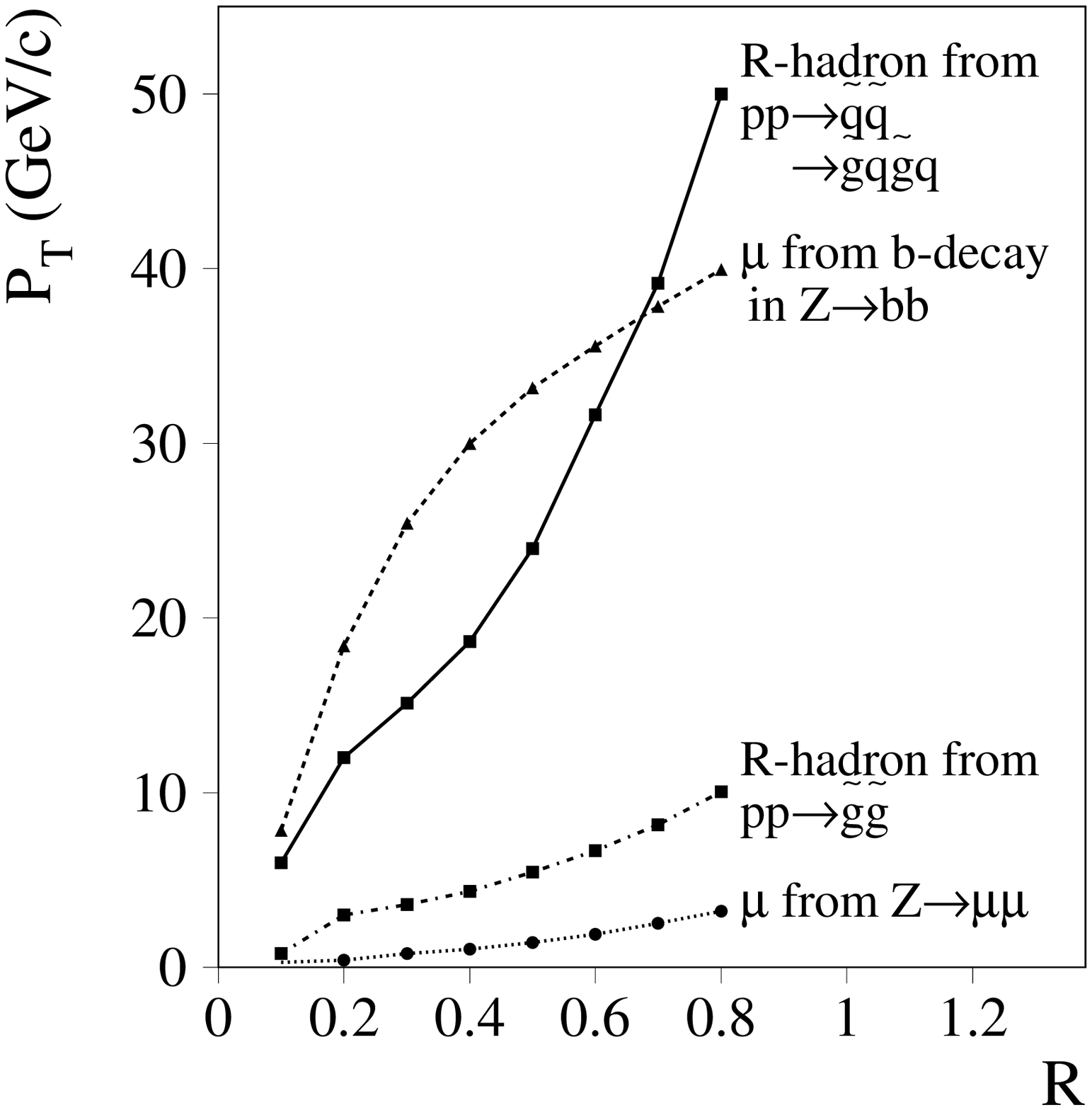,height=5.5cm,width=6cm}
\end{center}
\caption{\footnotesize A comparison of the jet structure between muons from $\rm p\rm p\rightarrow Z/\gamma\rightarrow\mu^+\mu^-$, muons from $\rm p\rm p\rightarrow Z/\gamma\rightarrow \rm b\bar{\rm b}$, and 300~\gevcc\ R-hadrons produced via \chanc~ and \chane. In this example the squark mass is set to 800~\gevcc. Left: mean value of the average number of particles $N$ in a cone around the muon direction as function of the cone radius $\rm R$. Middle: mean value of the particle density $N_{density}$ around the muon direction as function of the cone radius $\rm R$. Right: mean value of the transverse momentum, $p_T$, in a cone around the muon or R-hadron direction as function of the cone radius $\rm R$.   \label{fig:isolation}}
\end{figure}

%\subsubsection{Isolation}
Isolation may have an importance for the trigger efficiency for R-hadrons, as ATLAS plans to employ some form of isolation criteria to control the final trigger rate. Since we expect muons to be the dominant background source for R-hadrons, the jet structure around the gluino is investigated at event generator level and compared to that of isolated muons and non-isolated muons produced via b-quark decay. For this study, 50000 $\rm p\rm p\rightarrow Z/\gamma\rightarrow\mu^+\mu^-$ events and 50000 $\rm p\rm p\rightarrow Z/\gamma\rightarrow \rm b\bar{\rm b}$ events, with $b$ decaying weakly into muons, have been generated, and compared to R-hadrons of mass 300~\gevcc, produced via \chanc~ and \chane. For the latter process the mass of the squark must be larger than the mass of the gluino (LSP) and should be below $\sim$1 TeV in standard supersymmetry scenarios. In this example a squark mass of 800~\gevcc\ is used. In all cases the $\hat{p}_T$ of the hard $2\rightarrow 2$ process was required to be more than 50~GeV/c. Figure~\ref{fig:isolation} shows the distribution of the number of particles, the particles density, and the total $p_T$ of the particles in a cone of size $\rm R=\sqrt{\Delta\eta^2+\Delta\phi^2}$ around the muon or R-hadron, excluding the muon or R-hadron itself (and neutrinos). As can be seen, the typical R-hadron will not be perfectly isolated and some loss from an isolation criteria is to be expected.

\section{Event simulation}
\subsection{R-hadron event simulation}\label{sec:atlassignal}
There are no models available which propose definite gluino LSP masses following from GUT scale assumptions. Thus, the gluino LSP mass at the electroweak scale is a free parameter. Below we concentrate entirely on the channel~\chancc. Since the cross section for this channel depends only on the gluino mass, none of the masses of the other supersymmetric particles is relevant, as long as they are heavier than the gluino to assure non-decaying gluinos. 
The gluino masses investigated are $m_{\tilde{\rm g}}=100, 300, 600, 900, 1100, 1300, 1500, 1700$ and 1900~\gevcc\ and events were generated with PYTHIA~\cite{torbjorn}. 

The ATLAS fast simulation framework is used to simulate R-hadrons in the detector and perform the
physics analysis. The framework is extended to account for R-hadrons, using parameterizations of the momentum resolution of inner detector, the response in the calorimeter, and the response in the muon spectrometer. The parameterizations come from studying signatures in the different ATLAS sub-detectors for singly charged R-hadrons of various masses (100~\gevcc\ to 1900~\gevcc), transverse momenta and pseudo-rapidity ($\eta$). These have been fully simulated using the ATLAS detector simulation framework based on GEANT3~\cite{geantmanual}, and subsequently reconstructed with the ATLAS C++ reconstruction framework. A detailed discussion of the GEANT3~\cite{geantmanual} implementation of the model used can be found in Refs.~\cite{aafke,aafkethesis}. For the resolution of the inner detector momentum, the parameterization followed quite well that of muons in ATLAS~\cite{atlastdr},
%but we have modified the contribution due to multiple scattering (proportional to 1/($\beta$p)) in order to take into account the R-hadron velocities being smaller than the speed of light.
apart from an additional contribution due to multiple scattering (proportional to 1/($\beta$p)), especially important for slow R-hadrons, that was included.

 Typically, the momentum resolution, $\sigma \big(1/p_T\big)$, is of the order of 0.005. Details on the response of the calorimeters and the muon chambers, forming the base for the parameterizations, are described in Section~\ref{sigtrig}. Further information on the parameterizations can be found in Ref.~\cite{aafkethesis}.

\subsection{Background simulation}\label{sec:atbg}
QCD processes with their high cross section present a potential background problem for studies of hadronic particles. Fortunately, the cross section for QCD events drops rapidly with increasing transverse momentum scale in the hard process ($\hat{p}_T$), and the $p_T$ of the hadronized particles is further suppressed by the soft fragmentation function of light partons. Moreover, the long lifetime of pions and kaons effectively prevents them from decaying into muons before they get absorbed in the calorimeter. However, the weak decays of high $p_T$ b-quarks and t-quarks will produce high $p_T$ muons, and thus present a serious background. For the QCD background, separate samples have been generated for light quarks (denoted by QCD), b-quarks (denoted by b$\bar{\rm b}$), and t-quarks (denoted by t$\bar{\rm t}$). QCD b$\bar{\rm b}$ and t$\bar{\rm t}$ events have been generated in logarithmic $\hat{p}_T$ bins. The number of generated events in each $\hat{p}_T$ bin is assured to be larger than that corresponding to $1$ fb$^{-1}$. The only exception to this rule is in the region of very small $\hat{p}_T$, with very large cross sections, but out of several millions of generated events, no events pass the trigger cuts. Generating a number of events corresponding to 1 fb$^{-1}$ is beyond the scope of this work, and would only be relevant for a QCD trigger rate study.

Also, events with single Z and W production as well as events where two gauge bosons (denoted by WW/WZ/ZZ) form a potential background source. Such events may result in muons, and these events are likely to pass the trigger criteria. The $p_T$ value for most muons produced from Z and W is of the order of $M_Z/2$, and thus on average smaller than those of R-hadrons. Thus, we expect much of this background to be effectively removed by a $p_T$ cut on muons. The generation of Z, W, and diboson events is also done in bins. The width of the bins is optimized according to the cross section, and the number of generated events in each $\hat{p}_T$ bin is assured to be larger than that corresponding to $1$ fb$^{-1}$. 

\section{R-hadron triggers}\label{sigtrig}
%Due to their slow movement, yet high momentum, and their composite colored structure, with one constituent being ultra heavy, R-hadrons posses very distinct features when they traverse ordinary matter. The typical signatures for the detection of single heavy stable hadrons are
%\begin{itemize}
%\item High transverse momentum for charged hadrons.
%\item Large ionization in the tracking system, in case the R-hadrons are charged and slow.
%\item A characteristic pattern of energy deposition in the calorimeters.
%\item A large time-of-flight, measurable with the muon chambers.
%\end{itemize}

%\subsection{R-hadron triggers}\label{sigtrig}
Prior to selection, events are required to pass a set of trigger requirements. With their special signatures, R-hadrons may fall outside the standard trigger criteria in ATLAS~\cite{daqtdr1,daqtdr}. In the following we discuss the most relevant ATLAS trigger criteria for R-hadron events, as summarized in Table~\ref{tab:trigtab}. The muon triggers turn out to be the most promising, although for very high masses, the jet triggers have a sizable contribution.

The ATLAS trigger efficiencies are estimated by making a pseudo-trigger, which includes the trigger menus as defined in Table~\ref{tab:trigtab}. The pseudo-trigger distinguishes between a 'first level trigger', where only calorimeter information and the muon triggers are used, and a 'high level trigger', using all possible information. 
%The $1\mu 6$ trigger threshold was introduced in Ref~\cite{daqtdr1}, but later replaced by the 1$\mu$20 threshold in Ref~\cite{daqtdr} due to the higher (doubled) expected initial luminosity at the LHC startup. The local threshold (low level hardware) for {\em individual} muons in the ATLAS muon-system is still $p_T>6$~\gevc. This has no impact however on the R-hadron signal.%

\begin{table}[htb]
\begin{center}
\begin{tabular}{|l|l|}
\hline
Menu & Requirement \\
\hline
\hline
$Ex70j70$ & $E_T^{miss}>70$~GeV and a jet with $E_T>70$~GeV.\\
$j400$ & Jet  $E_T>400$~GeV.\\
$2j350$ & Two jets, each of which has $E_T>350$~GeV.\\
$Ex200$ & $E_T^{miss}>200$~GeV\\\hline
$1\mu 6$ & One muon, which has $p_T>6$~\gevc\footnotemark \\
$2\mu10$ & Two muons, each of which has $p_T>10$ \gevc.\\
% $1\mu20$ & One muon with $p_T>20$ \gevc.\\
\hline
\end{tabular}
\end{center}
\caption{\footnotesize Trigger menus and thresholds used in this study, they represent a subset of the full ATLAS trigger menus as originally defined in Ref.~\cite{daqtdr1} and later revised in Ref.~\cite{daqtdr}.\label{tab:trigtab}}
\end{table}

\subsection{Jet/$E_T^{miss}$ triggers}
For the jet/$E_T^{miss}$ triggers, it is difficult to estimate the efficiency precisely without dedicated full event simulation, because it depends on the details of the specific trigger reconstruction algorithms. Two aspects must be kept in mind when studying the interactions of R-hadrons in the calorimeter. First of all, an R-hadron passing the ATLAS calorimeters will undergo repeated charge flipping in subsequent interactions ($\sim$12) through the calorimeter. Hence, the results presented here are independent of the initial charge of the R-hadron for the momentum range studied. Second, an R-meson will convert to an R-baryon due to repeated nuclear interactions, because this is kinematically favorable~\cite{aafke}. Once the R-hadron is baryon, the nuclear cross section is 3/2 larger~\cite{aafke} and thus energy losses due to nuclear scattering increase.
\footnotetext{Introduced in Ref~\cite{daqtdr1}, but replaced in Ref~\cite{daqtdr} due to the higher (doubled) expected initial luminosity at the LHC startup. The local threshold for {\em individual} muons in the ATLAS muon-system is still $p_T>6$~\gevc. This has no impact on the R-hadron signal.}

\begin{figure}[bht]
\begin{picture}(0,70)
\put(15,0){\epsfxsize70mm\epsfbox{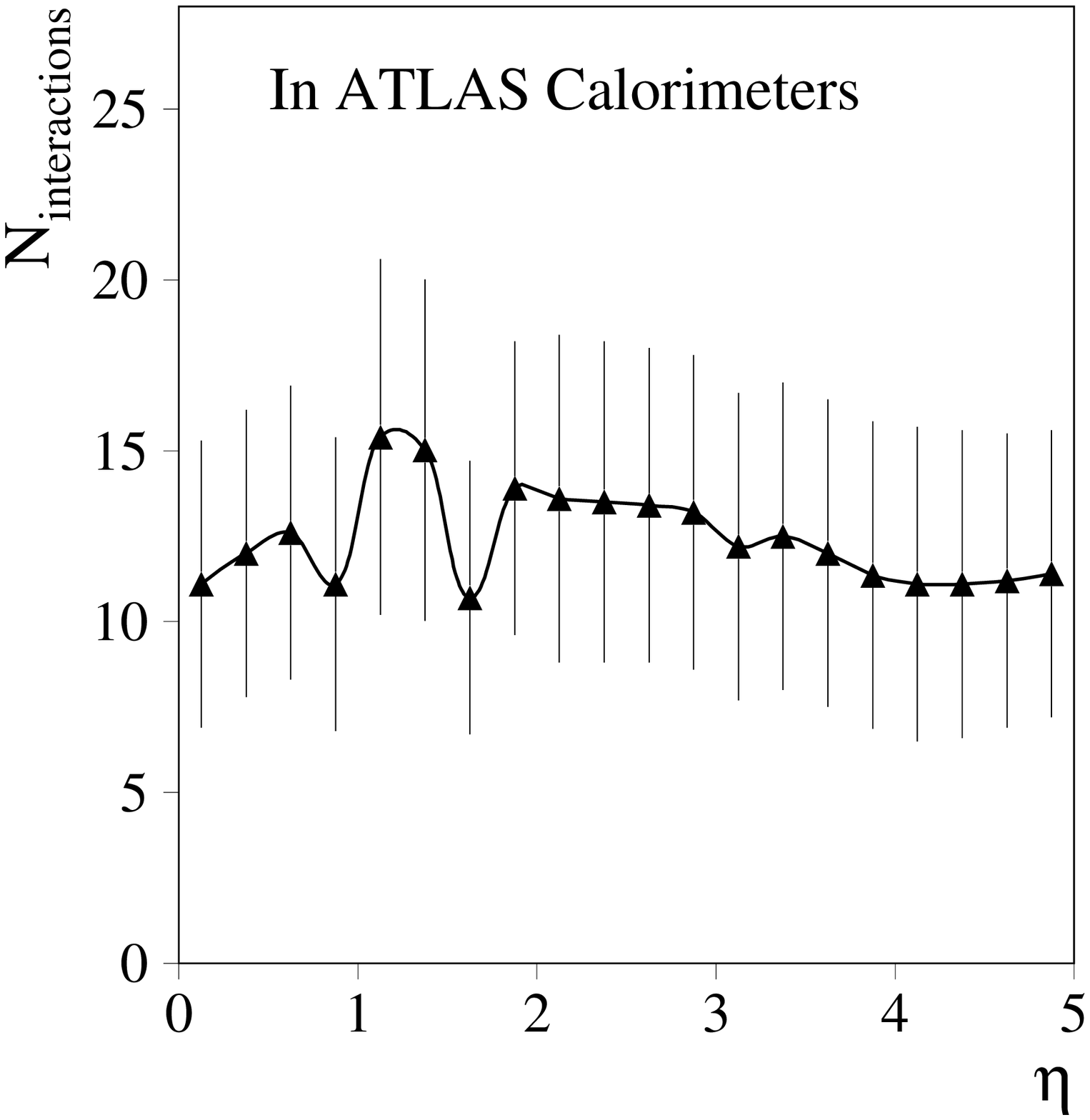}}
\put(63,49){(a)}
\put(85,0){\epsfxsize70mm\epsfbox{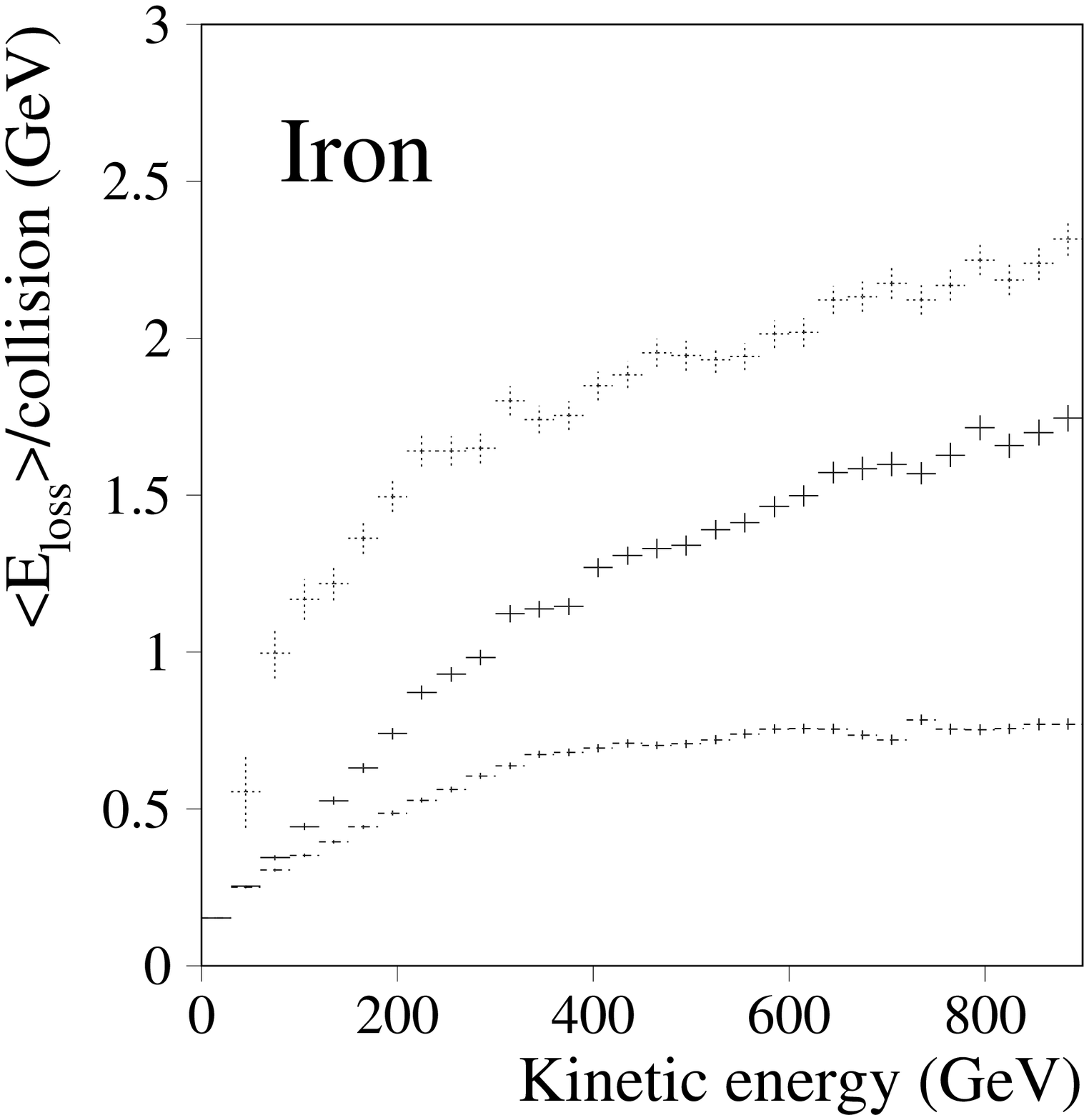}}
\put(133,49){(b)}
\end{picture}
\caption[ ]
{\protect\footnotesize (a) The average number of interactions and its
  spread at the exit of the ATLAS calorimeter system as function of $\eta$, for a R-hadron punching through
  the ATLAS calorimeters. These results are for
  full simulation with the ATLAS detector simulation framework based
  on GEANT3 and reconstruction with the ATLAS C++ reconstruction framework. (b) The average energy loss per interaction in iron for an R-hadron with a mass of 300~\gevcc. The lower and upper curves correspond to the energy loss enabling only $2 \rightarrow 2$ and $2 \rightarrow 3$ scattering, respectively, and the middle histogram represent the combination~\cite{aafke}.\label{fig:antint}} 
\end{figure}

The energy deposit and the possible stopping of R-hadrons in the calorimeters depend on the number of interactions an R-hadron undergoes when passing the calorimeters and the actual energy loss per interaction. 
In Fig.~\ref{fig:antint}a, the number of interactions for an R-hadron traversing the ATLAS calorimeters is displayed as a function of $\eta$. The typical energy loss per interaction in iron for an R-hadron with a mass of 300~\gevcc\ is shown in Fig.~\ref{fig:antint}b.
As can be seen from this plot, the energy loss for an R-hadron tends to level off for very high energies. For an R-hadron which punches through the calorimeter and passes the support structure to arrive at the first muon-station the number of interactions is about 15\% higher.

\begin{figure}[t!]
\begin{center}
\epsfig{file=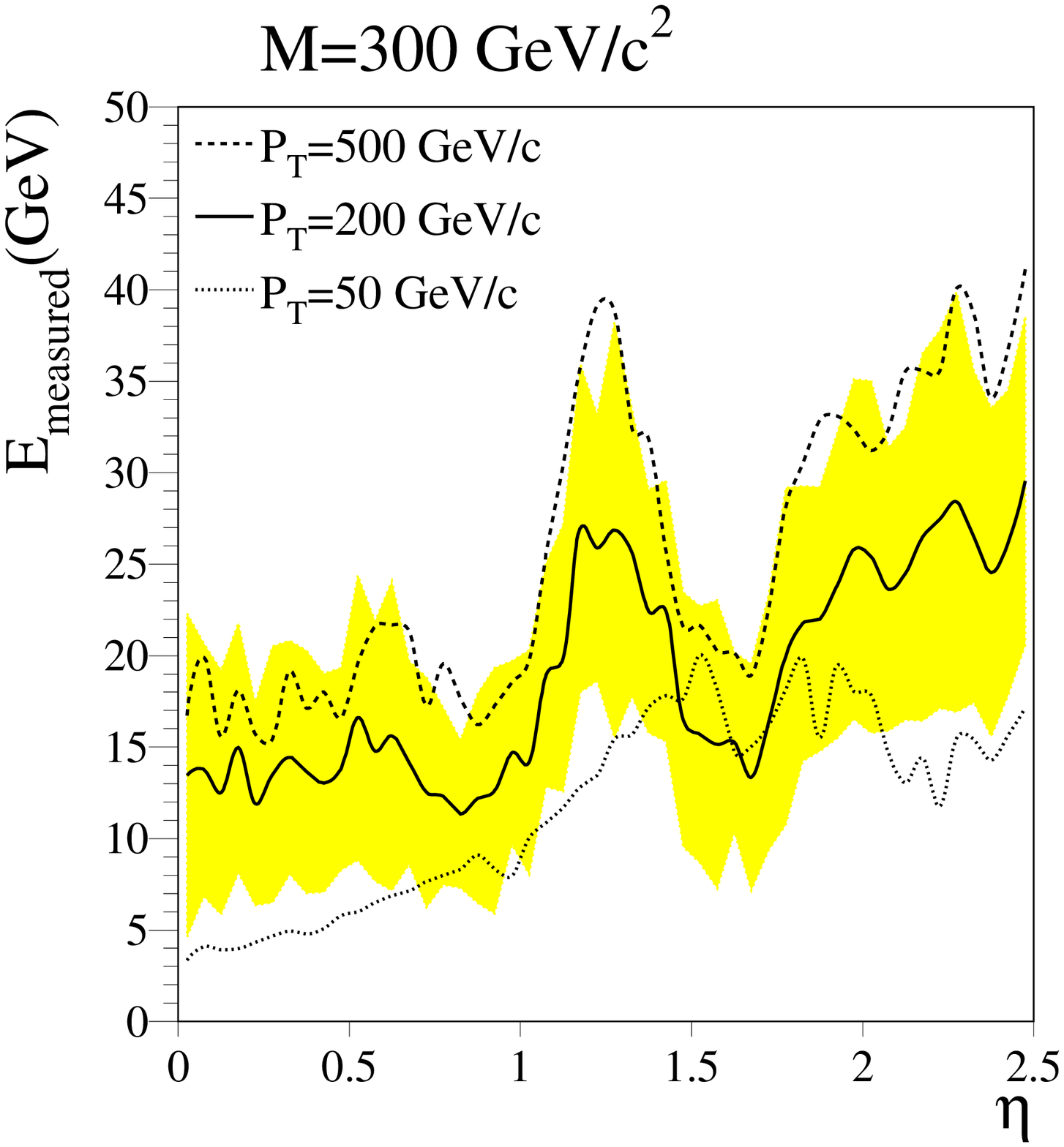,height=6.5cm,width=6.5cm}\epsfig{file=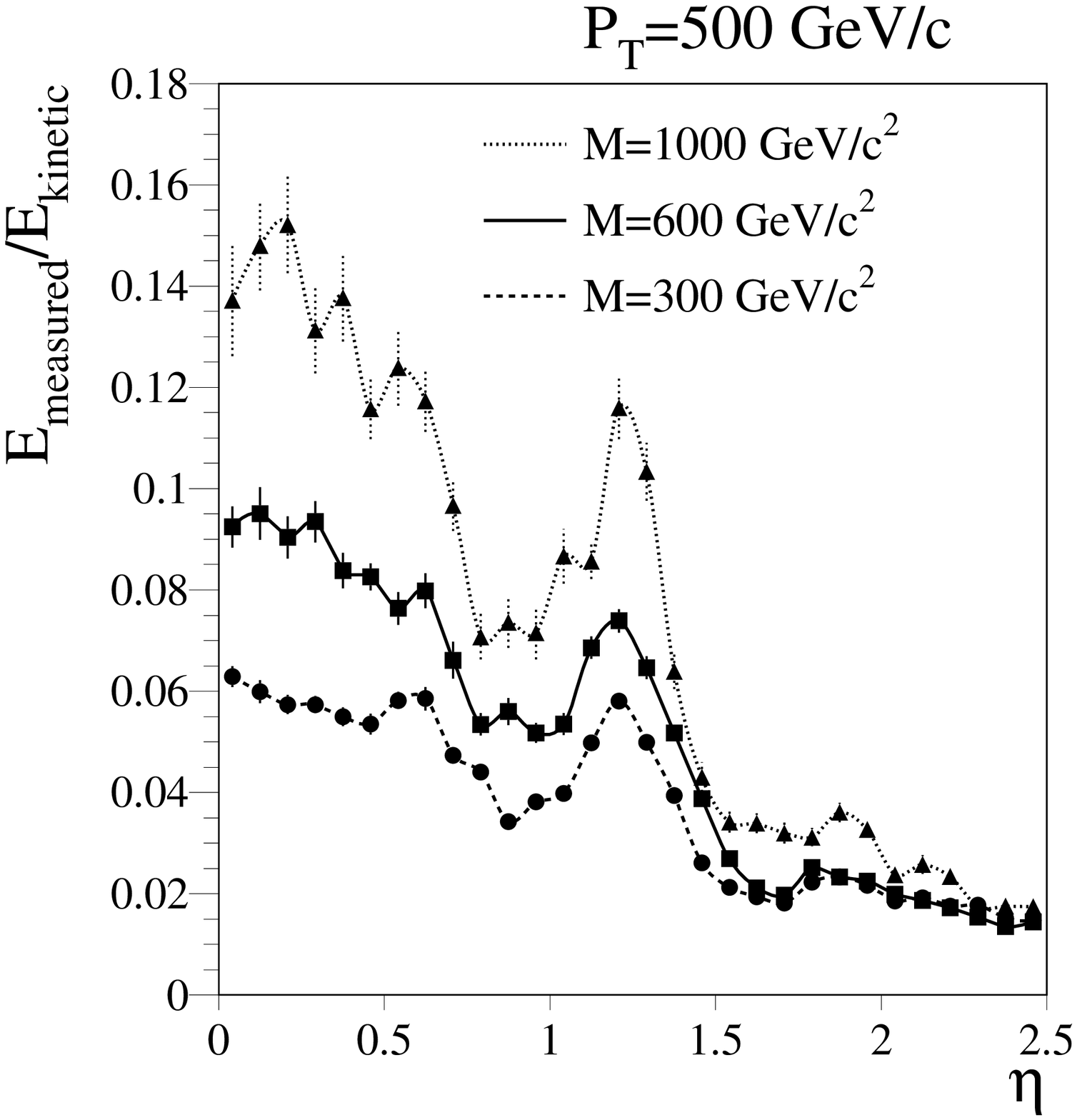,height=6.5cm,width=6.5cm}\end{center}
\caption{\protect\footnotesize Energy loss for punch-though
  R-hadrons. On the left-hand side the absolute energy loss is shown
  as function of $\eta$ for different transverse momenta. The shaded
  area represents the spread in the energy deposit for R-hadrons with
  mass 300~\gevcc\ and $p_T$ of 500~\gevc. On the right-hand side the
  relative energy deposition is plotted for R-hadron masses of 300,
  600 and 1000~\gevcc\ with a transverse momentum of 500~\gevc. Results are obtained by full simulation with the ATLAS
  detector simulation framework based on GEANT3 and
  reconstruction with the ATLAS C++ reconstruction framework.\label{fig:en}}
\end{figure}

The total energy deposit measured in the ATLAS calorimeters is displayed as function of $\eta$ for various R-hadrons masses and transverse momenta in Fig.~\ref{fig:en}. This figure shows that the relative energy loss drops with higher kinetic energy.

The first level jet/$E_T^{miss}$ triggers are based on a quick analysis of the regions of interest (ROI's) 
%\paragraph{Impact on Trigger}
in the calorimeters and the muon system, while at higher trigger levels this information is combined with tracking information~\cite{daqtdr}. Thus, for R-hadrons, it is crucial to understand the interplay between the information accessible at the different trigger levels: for example, it could happen that one charged and one neutral R-hadron are produced in the primary interaction. Depending on the details of the trigger algorithm and whether a charged R-hadron reaches the muon chamber, this could result in a large $E_T^{miss}$.  However, no such imbalance is present if considering the calorimeters alone, and it is therefore possible that such an event does not pass the first level calorimeter based $E_T^{miss}$ trigger. The same situation may happen for the calorimeter jet-based triggers: an R-hadron energy deposit rarely will give rise to a very high energetic jet in the calorimeters, but including the possible inner detector track information for charged R-hadrons in the final event reconstruction will result in a high $p_T$ jet.

\begin{table}%[t]
\begin{center}
\begin{tabular}{|c|c|c|c|c|c|c|}
\hline
\multicolumn{2}{|c|}{Trigger} & \multicolumn{5}{c|}{Mass (GeV/$c^2$)}\\ \cline{3-7}
\multicolumn{2}{|c|}{} & 100 & 300 & 600 &1100&1500\\ \hline
$Ex70j70$ & \begin{tabular}{l}Inclusive\\ Exclusive\end{tabular} & \begin{tabular}{c}6\%\\1\%\end{tabular} & \begin{tabular}{c}23\%\\5\%\end{tabular} & \begin{tabular}{c}36\%\\10\%\end{tabular} & \begin{tabular}{c}47\%\\18\%\end{tabular} & \begin{tabular}{c}52\%\\26\%\end{tabular}\\  \hline
$j400$    & \begin{tabular}{l}Inclusive\\ Exclusive\end{tabular} & \begin{tabular}{c}0\%\\0\%\end{tabular} & \begin{tabular}{c}1\%\\0\%\end{tabular} & \begin{tabular}{c}5\%\\1\%\end{tabular} & \begin{tabular}{c}9\%\\3\%\end{tabular} & \begin{tabular}{c}11\%\\5\%\end{tabular}\\  \hline
$2j350$   & \begin{tabular}{l}Inclusive\\ Exclusive\end{tabular} &  \begin{tabular}{c}0\%\\0\%\end{tabular} & \begin{tabular}{c}0\%\\0\%\end{tabular} &  \begin{tabular}{c}0\%\\0\%\end{tabular} & \begin{tabular}{c}1\%\\0\%\end{tabular} &  \begin{tabular}{c}1\%\\0\%\end{tabular}\\  \hline
$Ex200$   & \begin{tabular}{l}Inclusive\\ Exclusive\end{tabular} &  \begin{tabular}{c}0\%\\0\%\end{tabular} & \begin{tabular}{c}5\%\\1\%\end{tabular} & \begin{tabular}{c}13\%\\2\%\end{tabular} & \begin{tabular}{c}23\%\\7\%\end{tabular} & \begin{tabular}{c}27\%\\12\%\end{tabular}\\
\hline
\end{tabular}
\end{center}\caption{\protect\footnotesize An estimate of the final jet/$E_T^{miss}$ trigger
  efficiencies, after first and high level trigger, for \chanc~events at low luminosity. These results are based on the pseudo trigger
  simulation of R-hadron events in the ATLAS detector. The exclusive trigger efficiencies are for events passing a specific jet/$E_T^{miss}$ trigger and not the muon triggers.\label{tab:efftrigcal}}
\end{table}

Table~\ref{tab:efftrigcal} gives the final trigger efficiencies, after the first and high level triggers, for the jet/$E_T^{miss}$ trigger menus listed in Table~\ref{tab:trigtab}. As seen from Table~\ref{tab:efftrigcal}, an R-hadron event is most likely to pass the trigger when the energy deposit in the calorimeters results in a combined moderate energy jet and $E_T^{miss}$ trigger signal. Due to the moderate energy loss of an R-hadron in the calorimeters, combined with the high jet energy trigger threshold, the jet triggers only start to contribute for very high R-hadron masses. In particular, the efficiency of the two-jet trigger is negligible. The $E_T^{miss}$ trigger is also only effective for higher R-hadron masses. This is expected, since the typical momentum scales of R-hadron events follow the involved R-hadron masses, as discussed in Section~\ref{sec:topology}. Finally, it is seen that more than 50\%\ of the R-hadron events passing a jet/$E_T^{miss}$ trigger \emph{also} pass a muon trigger.

\subsection{Muon trigger}\label{sec:muontrigger}
The possibility to use the muon trigger for the detection of slow
charged particles, not interacting hadronically, has already been
investigated for the central region of the ATLAS
detector~\cite{nisati}. The trigger efficiencies were estimated by
requiring a coincidence between the two involved trigger stations
within a time window of 18~ns. For low luminosity, the efficiency of
the trigger was found to be as high as 90\% for slow particles with
$\beta=0.2$. However, such slow particles would arrive
at the muon stations 125~ns after the bunch crossing, or 100~ns after
an ordinary muon would arrive. Thus, when
the muon trigger condition is met, and data-taking started, the information of
an event which came four bunch crossings later is recorded, and the
relevant information of the slow particle from the inner detector 
and calorimeters is probably already lost. To which extent the
relevant information is available, depends on the velocity of the
R-hadron and the detailed timing of the signal in the different
subsystems.

In the following, it is assumed that a priori about 75\%\ of all R-hadrons shows up as charged in the muon system irrespective of their original charge and type (R-meson or R-baryon) when they were formed in the primary interaction. This is due to the many repeated nuclear interactions upstream to the muon system, after which an R-hadron independently of it being an R-meson or R-baryon will end up as either a charged (3 states) or neutral (1 state) R-baryon at the muon chambers. This effect is included in the final trigger estimates presented in Table~\ref{tab:trigeff}.

%\paragraph{Estimate of muon trigger efficiency based on full simulation}\label{mu:cond}

The ATLAS muon trigger has 2 different configurations using different muon stations, depending on the instantaneous luminosity~\cite{atlastdr}. For low luminosity the trigger requires a coincidence between the two barrel RPC's (Resistive Plate Chambers) positioned at approximately 6.7 (1st RPC) and 7.5~m (2nd RPC) from the interaction point, or the two endcap TGC's (Thin Gas Chambers) located at approximately 14 (2nd TGC) and 14.5~m (last TGC). At high luminosity, a coincidence is required between different muon stations: the two barrel RPC's situated at approximately 6.7 (1st RPC) and 10~m (last RPC), or the two endcap TGC's placed at approximately 13 (1st TGC) and 14.5~m (last TGC). The efficiency of the muon trigger is studied by requiring in addition that the R-hadron is recorded in the correct event. 

The conditions of an R-hadron firing the muon trigger and being assigned to the right event are then the following
\begin{itemize} 
\item The R-hadron arriving at the muon stations must be charged and must have $\eta<2.5$.
\item The R-hadron has a coincidence in the two trigger stations within the temporal gate of the trigger stations, which is taken to be 18~ns~\cite{nisati} for both barrel and endcap trigger stations. For example, for the RPC's at $\theta=90^\circ$, this would correspond to $\beta>0.5$ for the high luminosity trigger stations and $\beta>0.14$ for the low luminosity trigger stations.
\begin{figure}[t!]
\begin{center}
\epsfig{file=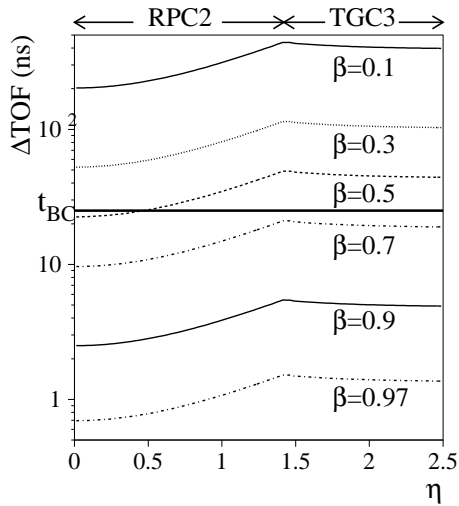,height=7cm,width=7cm}
\end{center}
\caption[]
{\protect \footnotesize The difference in time-of-flight $\Delta$TOF at RPC2 (7.5~m) and TGC3 (14.5~m) between an non-interacting particle traveling with velocity $\beta$ in a certain $\eta$-region, and a particle traveling with the speed of light, as function of $\eta$. The bunch crossing time $t_{BC}$ is the maximum difference allowed by the trigger.
\label{fig:delay1}}
\end{figure} 
\item The delay of a heavy particle compared to that of a muon must be
  less than 25~ns, so that the trigger is fired within the right
  event. This puts an upper limit on the maximum time-of-flight (TOF)
  of a particle at the trigger stations\footnote{The TOF information
    in GEANT is not available by default and has been introduced in
    the ATLAS detector simulation framework based on GEANT3 for this purpose.}. In Fig.~\ref{fig:delay1}, the difference in time-of-flight $\Delta\rm TOF$ between a non-interacting particle with $\beta<1$ and a particle with $\beta=1$ at RPC2 (7.5~m) and TGC3 (14.5~m) is shown. The latter stations are the last involved stations for a low luminosity muon trigger. Looking at the central detector region ($\eta=0$), we see that the particle must have $\beta\gtrsim 0.5$ in order to reach the last trigger station (7.5~m) in time. For a high luminosity trigger making use of the last trigger station in the barrel region (10 m), this limit would be $\beta>0.6$. In Fig.~\ref{fig:delay}, the delay at the last trigger stations for low luminosity is shown for fully simulated R-hadrons of mass 300~\gevcc\ of different $p_T$ values as function of $\eta$. Due to the increased energy loss per nuclear scattering, a light R-hadron is more delayed than a heavy one with the same starting velocity, when measured in the muon system, as shown in Fig.~\ref{muonbe}. 
\begin{figure}[t!]
\begin{center}
\epsfig{file=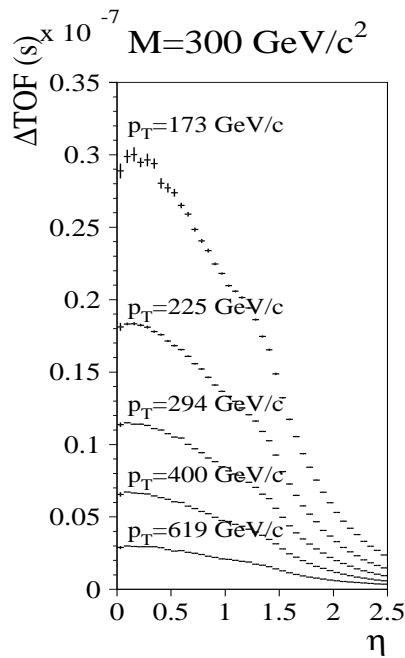,height=8.8cm,width=6cm}
\end{center}
\caption[]
{\protect \footnotesize The difference in time of flight $\Delta$TOF as function of $\eta$ at RPC2 and TGC3 between an R-hadron of mass 300~GeV/$c^2$ and a particle traveling with $\beta=1$. The displayed momenta correspond to $\beta$-values at $\eta=0$ of 0.5, 0.6, 0.7, 0.8, 0.9 and 0.98.
\label{fig:delay}}
\end{figure}

\item The deflection angle of the R-hadron in the magnetic field must
  be smaller ($p_T$ greater) than that of the corresponding muon
  (6 or 10~\gevc). Since R-hadrons are produced with high momenta, this condition is always satisfied. 
 \end{itemize}

The $1\mu 6$ and the  $2\mu 10$ trigger menus are studied by using fully GEANT 3 simulated single R-hadrons. The R-hadrons are then required to satisfy the above criteria and the resulting
trigger efficiencies are shown in Table~\ref{tab:efftrig} for \chanc. The shift towards lower values of $\beta$ for higher R-hadron masses has a clear negative impact on the efficiencies, in contrast to the jet/E$_T^{miss}$ triggers. The full simulation trigger studies presented here does not include the effect a possible isolation criteron. Depending on the initial trigger rates in ATLAS, some form of an overall isolation criteria can be introduced, although it is not present in the standalone muon system. As seen in Section~\ref{sec:topology}, the typical R-hadron is not produced isolated and using a simplified pseudo-trigger implementation of the criteria in Ref.~\cite{daqtdr} leads to a loss of about 50\%\ of the events fairly independent of the R-hadron mass. In contrast, an increase of trigger threshold for the muon triggers has no impact on the R-hadron signal.

\begin{table}[t!]
\begin{center}
\begin{tabular}{|c|c|c|c|c|c|}
\hline
Trigger & Mass & Luminosity &Efficiency (\%)& Maximum loss (\%)\\ 
& (GeV/$c^2$) & & \chanc & (Charge flip)\\
\hline
\hline
        & 100 & Low  & 61 & 3\\
        &     & High & 58 & 9\\
%1$\mu$6~\footnotemark & 300 & Low & 67 & 3\\
1$\mu$6 & 300 & Low & 67 & 3\\
        &     & High & 63 & 9\\
        & 600 & Low  & 60 & 3\\ 
        &     & High & 59 & 10\\ 
        & 1100 & Low & 55 & 4 \\ 
        &  & High & 52 & 12\\ 
        & 1500 & Low & 45 & 5 \\ 
        &  & High & 43 &13  \\ 
\hline
        & 100 & Low  & 28 & 11\\
        &     & High & 24 & 26\\
2$\mu$10& 300 & Low  & 33 & 12\\
        &     & High & 29 & 28\\
        & 600 & Low  & 34 & 13\\ 
        &     & High & 27 & 30\\ 
        & 1100 & Low & 21 & 13\\ 
        & & High & 17 & 31\\ 
        & 1500 & Low & 16 & 13\\ 
        &  & High & 13 & 32\\ 
\hline
\end{tabular}
\end{center}\caption{\protect\footnotesize An estimate of the trigger
  efficiency for \chanc~events. The fourth column shows the results for fully
  simulated R-hadrons in the ATLAS detector and requiring the conditions given in Section~\ref{sec:muontrigger}. The corresponding trigger efficiencies at low luminosity obtained from a fast pseudo trigger simulation agree within about 5\%. The fifth column indicates the potential maximum relative loss of events passing the trigger, due to nuclear interactions inside the muon system.\label{tab:efftrig}}
%\begin{tabular}{|c|c|c|c|c|c|}
%\hline
%Trigger & Mass & Luminosity &Efficiency (\%)&Maximum loss (\%)\\ 
%& &(GeV/$c^2$) & \chanc & (Charge flip)\\
%\hline
%\hline
%        & 100 & Low & 61  & 3.2\\
%        &     & High & 58 & 8.8\\
%1$\mu$6~\footnotemark & 300 & Low & 67 & 3.3\\
%        &     & High & 63 & 9.2\\
%        & 600 & Low & 60  & 3.1\\ 
%        &     & High & 59 & 10\\ 
%\hline
%        & 100 & Low & 28  & 11\\
%        &     & High & 24 & 26\\
%2$\mu$10& 300 & Low & 33  & 12\\
%        &     & High & 29 & 28\\
%        & 600 & Low & 34  & 13\\ 
%        &     & High & 27 & 30\\ 
%\hline
%\end{tabular}
%\end{center}\caption{\protect\footnotesize An estimate of the trigger
%  efficiency for \chanc~events. These results are based on fully
%  simulated charged R-mesons in the ATLAS detector and requiring the conditions given in Section~\ref{sec:muontrigger}. The last column indicates the potential maximum loss due to nuclear interactions inside the muon system.\label{tab:efftrig}}
\end{table}

In the calculations of the efficiencies of the muon trigger, all
charged R-hadrons reaching the muon chambers are assumed to produce
hits in {\em all} the trigger planes. Due to support structures and the presence
of the magnet, it is possible that additional nuclear reactions may
take place. For high $p_T$ muons, the probability to undergo a nuclear
interaction is very small, and the efficiency of a muon to fire the
trigger is very close to 100\%~\cite{atlastdr}. For an R-hadron,
however, the probability for nuclear interactions is considerably
larger. Since the support structures depend on $\eta$, the probability
has been evaluated with full simulation in five different $\eta$
regions.
\begin{table}[t!]
\begin{center}
\begin{tabular}{|l|c|c|c|c|}
\hline
 region & $\eta$ & 5-10 m (total)  & 6.7-7.5 m (LL) & 6.7-9.9 m (HL)\\
\hline
 & $\eta(0,0.5)$ & 30\% & 7\% & 17\% \\
 barrel:& $\eta(0.5,1.0)$ & 40\% & 10\% & 30\% \\
& $\eta(1.0,1.5)$ & 45\%& 7\% & 15\% \\
\hline
\hline
 region & $\eta$ & 7-15 m (total)  & 14.1-14.5 m (LL) & 12.9-14.5 m (HL)\\
\hline
 endcap &  $\eta(1.5,2.0)$ & 60\% & 2\% & 3\%\\
 & $\eta(2.0,2.5)$& 75\% & 2\% & 3\%\\
\hline
\end{tabular}
\end{center}
\caption{\footnotesize Fraction of R-hadrons undergoing one or more nuclear interactions in the muon system. The probabilities are displayed for the entire system, and for the region in between the two trigger stations at low (LL) and high (HL) luminosity.\label{tab:intmuon}}
\end{table}
The results, given in Table~\ref{tab:intmuon}, show that there is indeed a non-zero probability that an R-hadron interacts in the muon system and possibly converts into another R-hadron with a different charge. This aspect introduces an extra loss of efficiency for the muon triggers. The maximum loss, assuming that any interaction in the muon system implies a lost particle, is shown for the trigger in Table~\ref{tab:efftrig}. The numbers in Table~\ref{tab:efftrig} reflect a worst case scenario, as additional events where a neutral R-hadron at the entry of the muon-system becomes charged prior to arriving at the trigger stations, has not been included.

\subsection{Effects of time delays on the R-hadron measurement}
Although it is required that the R-hadrons are in time to be recorded within the right event, the event reconstruction may face additional problems as it is not a priori certain whether the full detector information of all the separate subsystems can still be used when reconstructing the complete event. Below we discuss the time delays in the different ATLAS subdetectors, starting from the detector systems closest to the interaction point and ending with the muon system.

For the pixels and silicon detectors, the delay in the central region is very small (maximum~1 and 2~ns for pixels and silicon, respectively), since their location is very close to the vertex. Second, for these detectors a hit is based on detecting a certain amount of charge~\cite{idtdr}. Slow R-hadrons are late, but would be heavily ionizing and certainly result in a hit. Third, we only make use of these detectors for momentum measurements, requiring only the hit positions, rather than the whole signal shape. Thus, the information of these subdetectors is reliable, if indeed we require the R-hadron to arrive at the muon chambers with a delay smaller than 25~ns. 

In the case of the TRT, the delay is slightly larger, maximally (for an R-hadron with velocity $\beta=0.5$) about 4~ns at the outer layers at 1.1 m. However, this delay is only a small fraction of the maximum drift time is of 40~ns, and it would correspond to a drift distance of only 200 $\mu$m. A good reconstruction of the signal is feasible. 

Heavily delayed R-hadrons imply additional complications for the jet/$E_T^{miss}$ triggers. In this case the R-hadron signal may not be suitable in time for the calorimeter (trigger) electronics to reconstruct the full pulse-shape of the signal, resulting in a lower reconstructed energy. Fortunately this is expected to have a small effect for two reasons. Firstly, the delay in arrival at the calorimeters is typically small (distance less than 4m), such that it is possible to adjust the delay timing to account for this if necessary. Secondly, for first level triggers based on $E_T^{miss}$, an R-hadron energy deposit which does not result in a ROI, will contribute to an even higher missing energy signal. For full reconstruction of the calorimeters, the time-interval over which data taking stretches in case of a positive trigger signal is roughly 125 and 175~ns for LAr and Tile calorimeters, respectively. Thus the signal can probably be fully reconstructed, although the resolution may be worse due to the time-shift w.r.t. a standard signal, for which the detector readout and timing have been optimized. 

For the muon system, the measurement by the
individual muon stations will not be affected by
delays, as the ATLAS muon system will run in a
continuous data taking model and with a long read
out time of about 500 ns~\cite{muontdr}.  Hence
measurement by the individual muon detectors are not
expected to be affected by time delays below 25 ns.
 However, the large overall dimension of the overall
muon system could possibly introduce time
incoherence effects when assigning a slow particle
measurement to the bunch crossing that was triggered on. In a worse case scenario of a large
delay, the most outer muon chambers could assign a
slow traversing charged particle to another bunch
crossing than the most inner chambers.
Although possibly important, studying this effect
requires a proper time simulation analysis and a realistic
description of the muon trigger at the time of
data-taking, and at the time of writing this is not
available in the ATLAS simulation.  For R-hadrons
which are successfully fully measured by the muon
detectors and assigned to the correct event, the
subsequent reconstruction is possibly affected. As
seen in Section~\ref{tofvariable}, the time delay will imply a
larger $\chi^2$ of the track fit. It is not expected to be a problem to relax the
software criteria to include the possibilty to
reconstruct particles with $\beta$ less than one.

\subsection{Trigger rates}
The pseudo trigger used in this study assumes the LHC to be running in the \emph{low luminosity mode}, because the cross sections for gluino production are anyway high, and the low luminosity muon trigger has a larger efficiency than the high luminosity trigger.
                                                                                                              
In Table~\ref{tab:trigeff} the numbers of events expected in 1 fb$^{-1}$ after triggering are summarized for different R-hadron masses and
background processes.

\begin{table}[t!]
\begin{center}
\begin{tabular}{|l|c|c|c|c|}
\hline
Sample & $\rm N_{\rm trig}(1\rm\ fb^{-1}$)\\
\hline
\hline
$M_{R}=$100 & $3.6\times 10^7$ \\ 
$M_{R}=$300 & $2.0\times 10^5$ \\
$M_{R}=$600 & $3.6\times 10^3$  \\
$M_{R}=$900 & $2.3\times 10^2$ \\
$M_{R}=$1100 & $4.9\times 10^1$ \\
$M_{R}=$1300 & $1.1\times 10^1$ \\
$M_{R}=$1500 & $3.2$ \\
$M_{R}=$1700 & $0.9$ \\
$M_{R}=$1900 & $0.3$ \\
\hline
QCD & 6.5$\times 10^8$ \\
$\rm b\bar{\rm{b}}$ &  $4.9\times 10^{8}$\\
W & $1.1\times 10^7$ \\
Z  & $5.3\times 10^6$ \\
diboson & $1.6\times 10^5$ \\
\hline
\hline
\end{tabular}
\end{center}
\caption{The number of triggered events at low luminosity running for R-hadrons of various masses and for the background sources for an integrated luminosity of 1~fb$^{-1}$. \label{tab:trigeff}}
\end{table}

\section{R-hadron event selection}~\label{sign}
\begin{figure}[t!]
\begin{center}
\epsfig{file=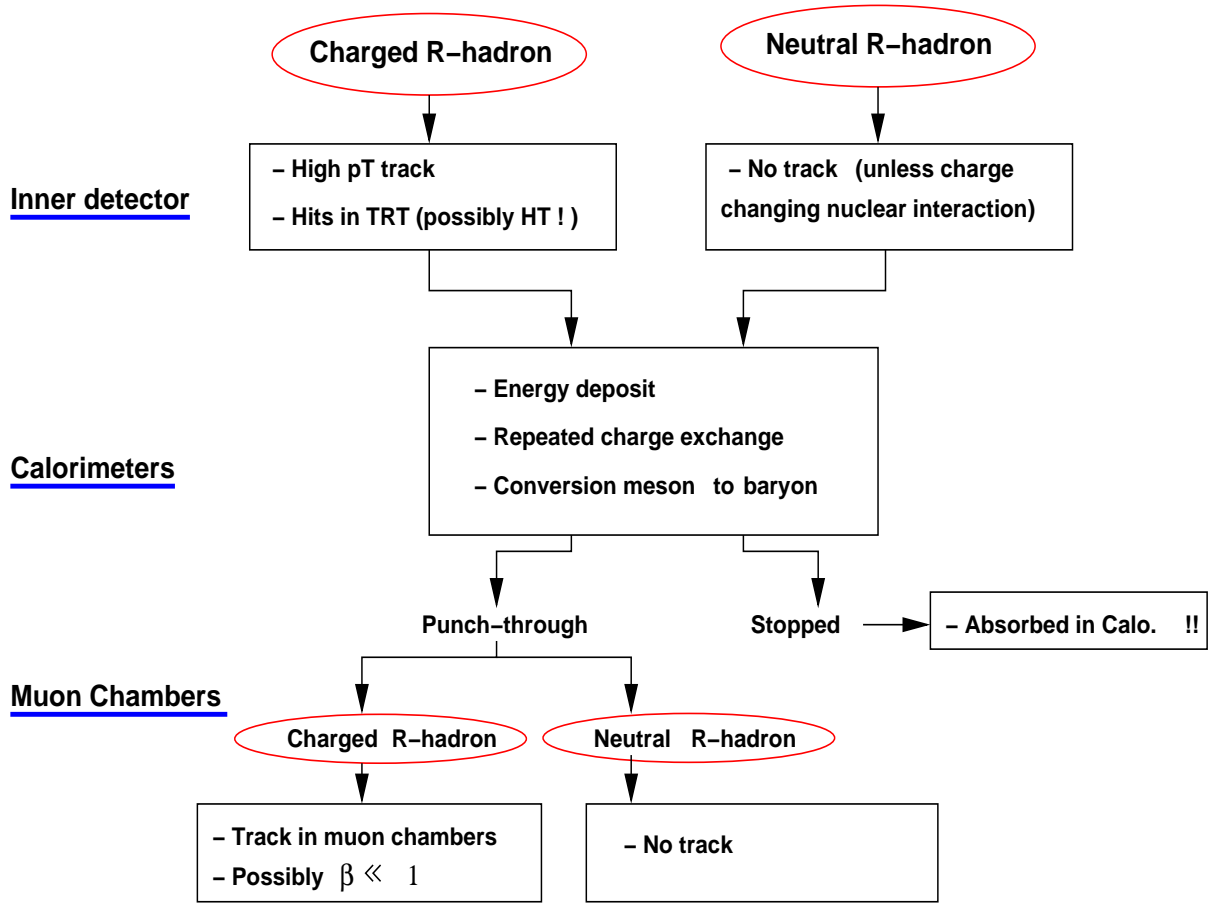,height=12cm,width=16cm}
\caption[ ]
{\protect\footnotesize Possible manifestations of an R-hadron in ATLAS. 
\label{rhadcomb}}
\end{center}
\end{figure}

A schematic overview of the R-hadron overall behavior in the ATLAS detector is shown in Fig.~\ref{rhadcomb}.

In the following two selection approaches have been investigated. First, a selection is performed on the basis of \emph{event topology} alone. This represents what can be done within the present analysis framework, i.e. standard event filter, and which does not require a very fine level of simulation of the response of the ATLAS detector. Second, another selection procedure is presented requiring a more detailed study of the possibilities to measure the \emph{time-of-flight} or velocity of the R-hadron candidate. The standard ATLAS reconstruction is used to reconstruct events containing R-hadrons. Hence, only for initially charged R-hadrons, a high $p_T$ charged track will be reconstructed in the inner detector. In contrast, all R-hadrons will produce relatively low energetic clusters in the calorimeter and R-hadrons arriving as charged at the muon system will be reconstructed by the muon system as a high $p_T$ muon assuming a $\beta$ of unity.

\subsection{Event topology selection variables}
\begin{figure}[t!]
\begin{center}
\hspace{-0.3cm}\epsfig{file=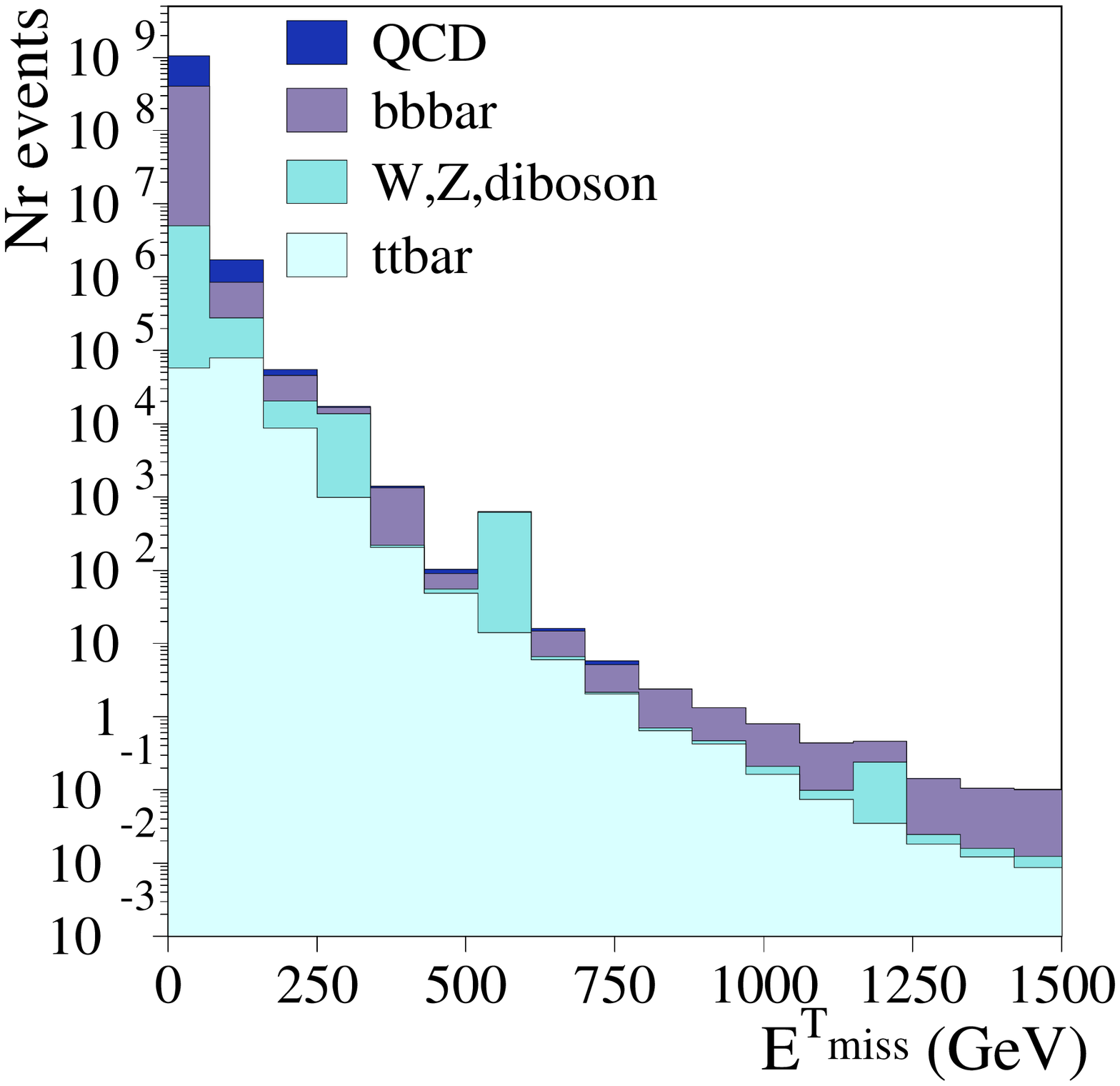,height=5.7cm,width=5.6cm}\hspace{-0.3cm}\epsfig{file=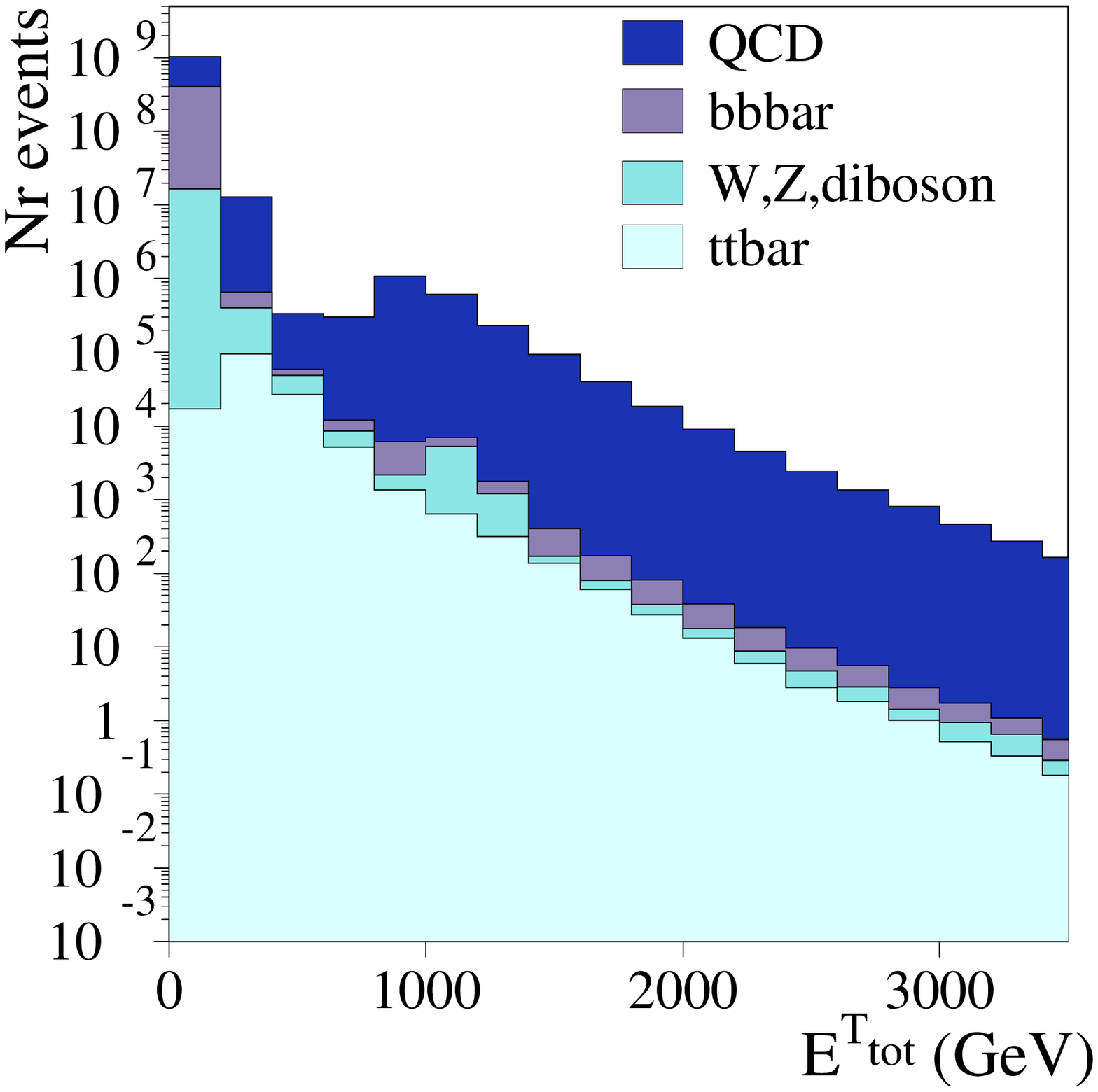,height=5.7cm,width=5.6cm}\hspace{-0.3cm}\epsfig{file=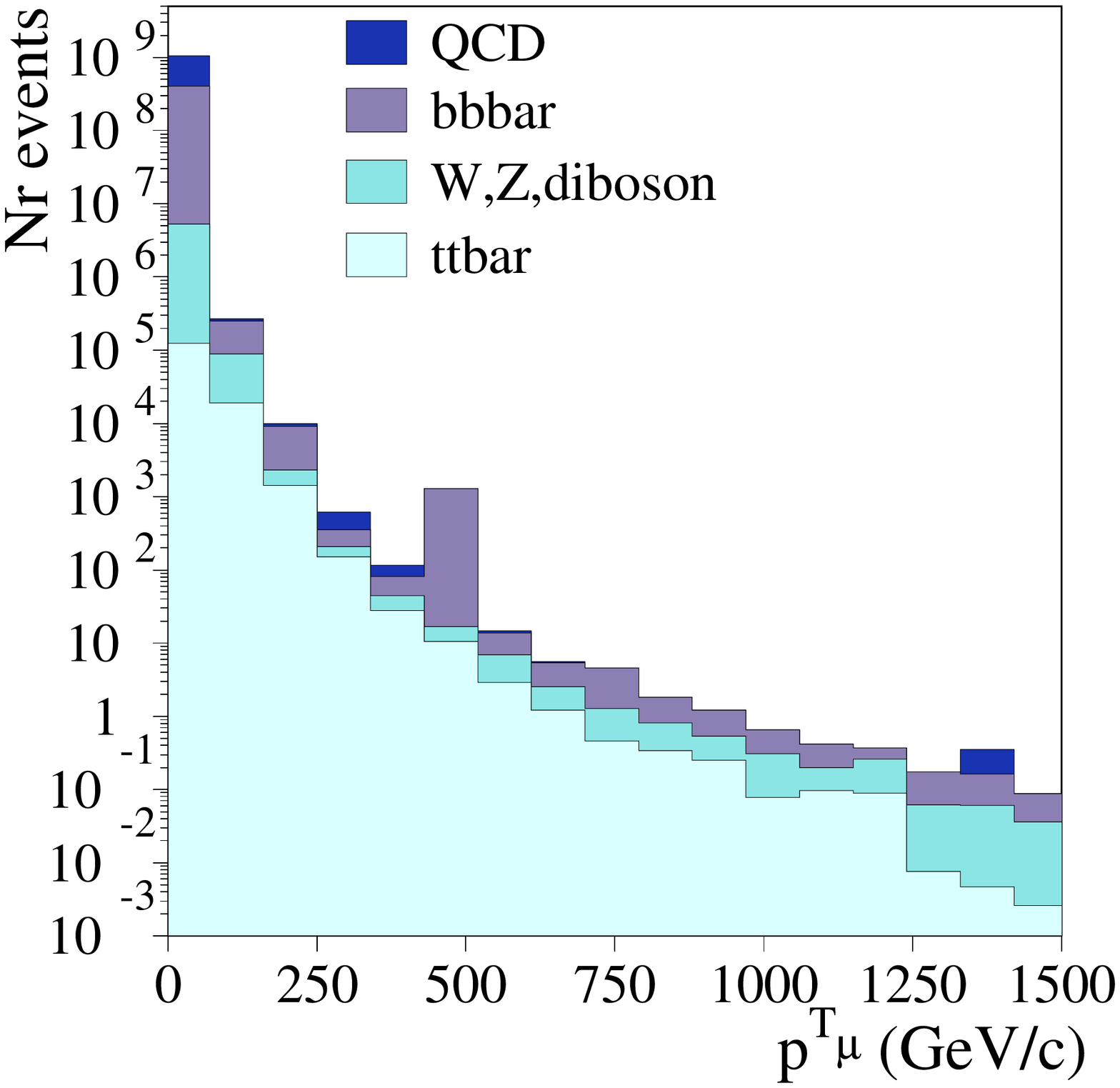,height=5.7cm,width=5.6cm}\\
\hspace{-0.3cm}\epsfig{file=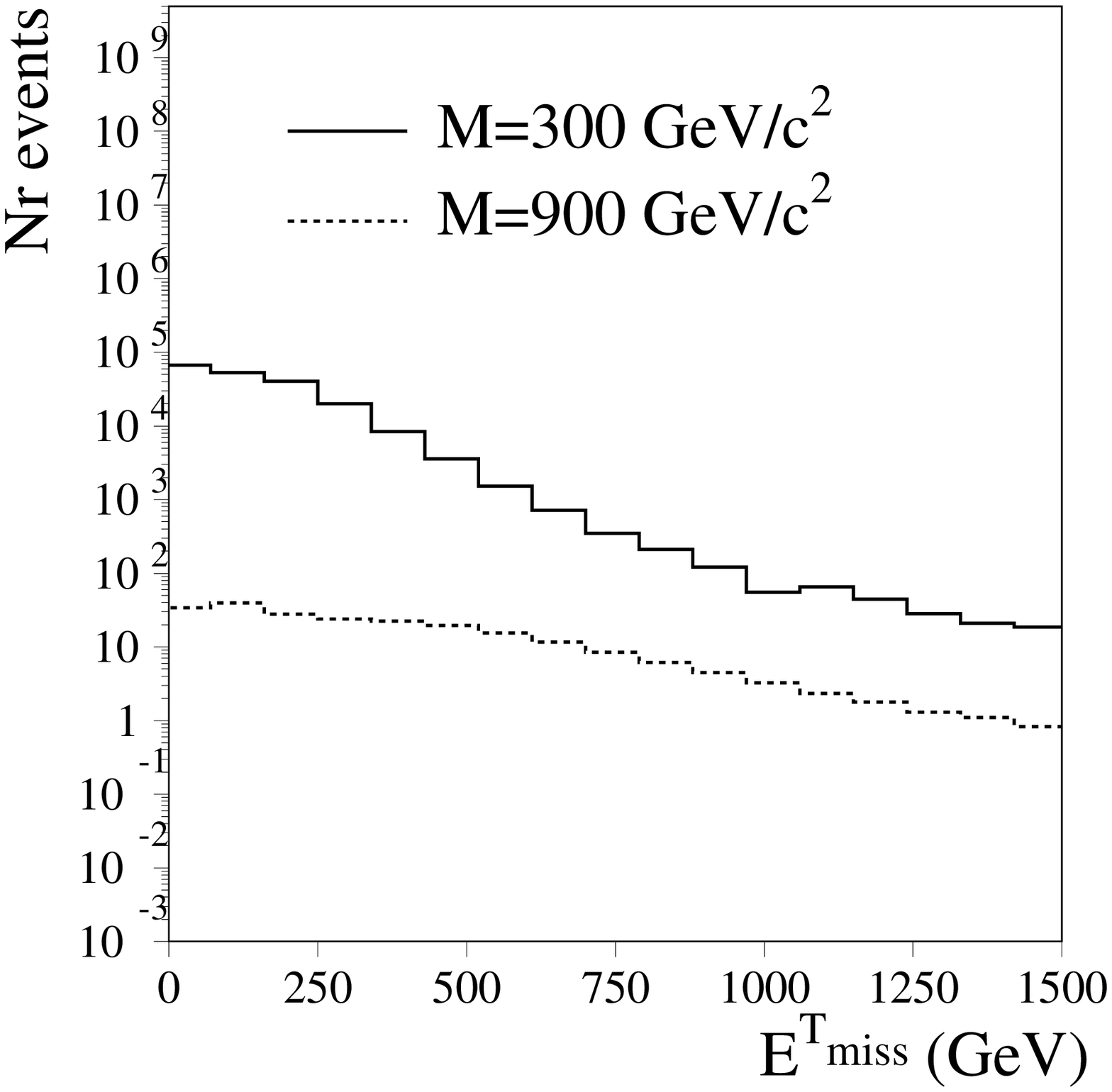,height=5.7cm,width=5.6cm}\hspace{-0.3cm}\epsfig{file=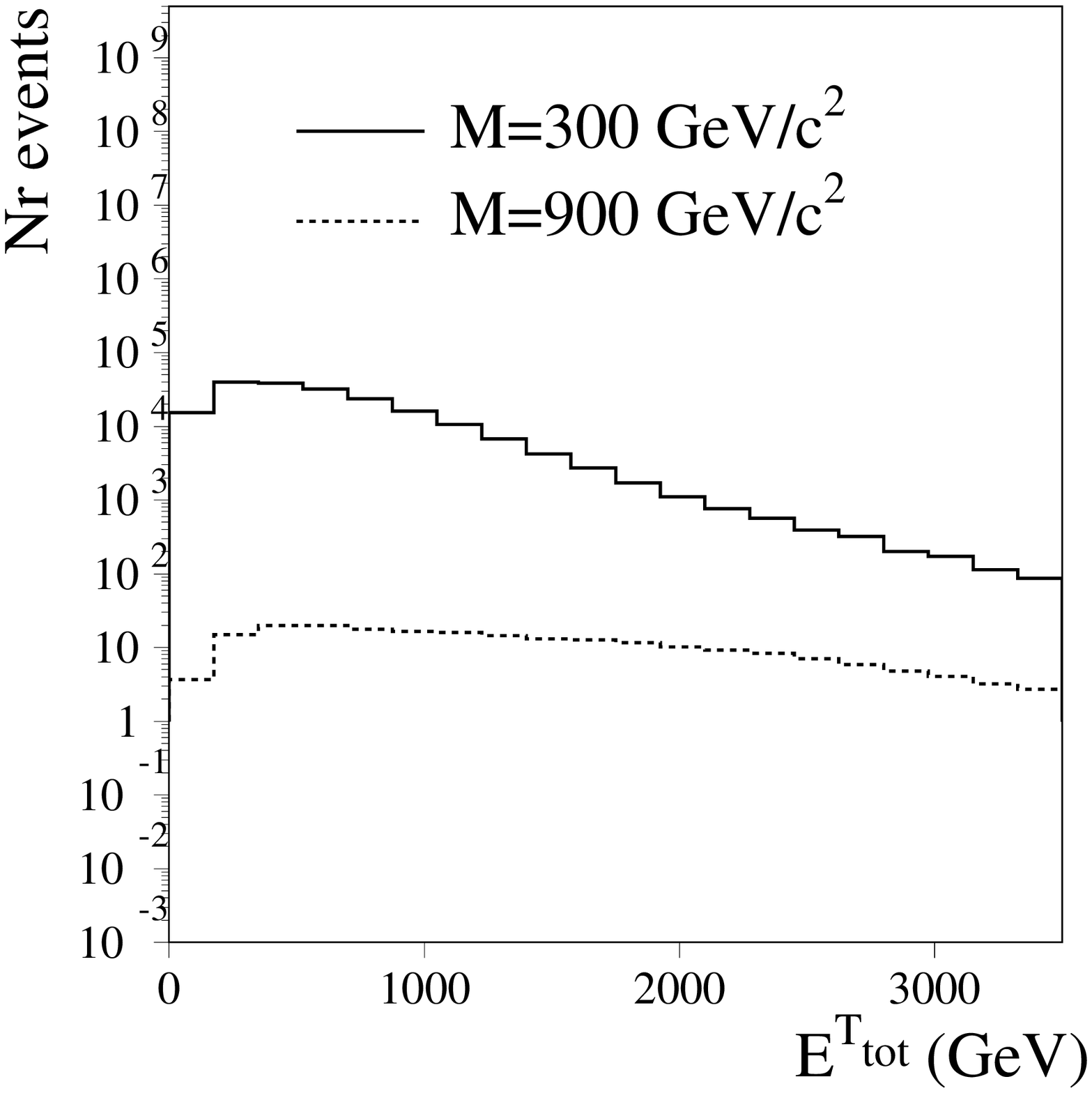,height=5.7cm,width=5.6cm}\hspace{-0.3cm}\epsfig{file=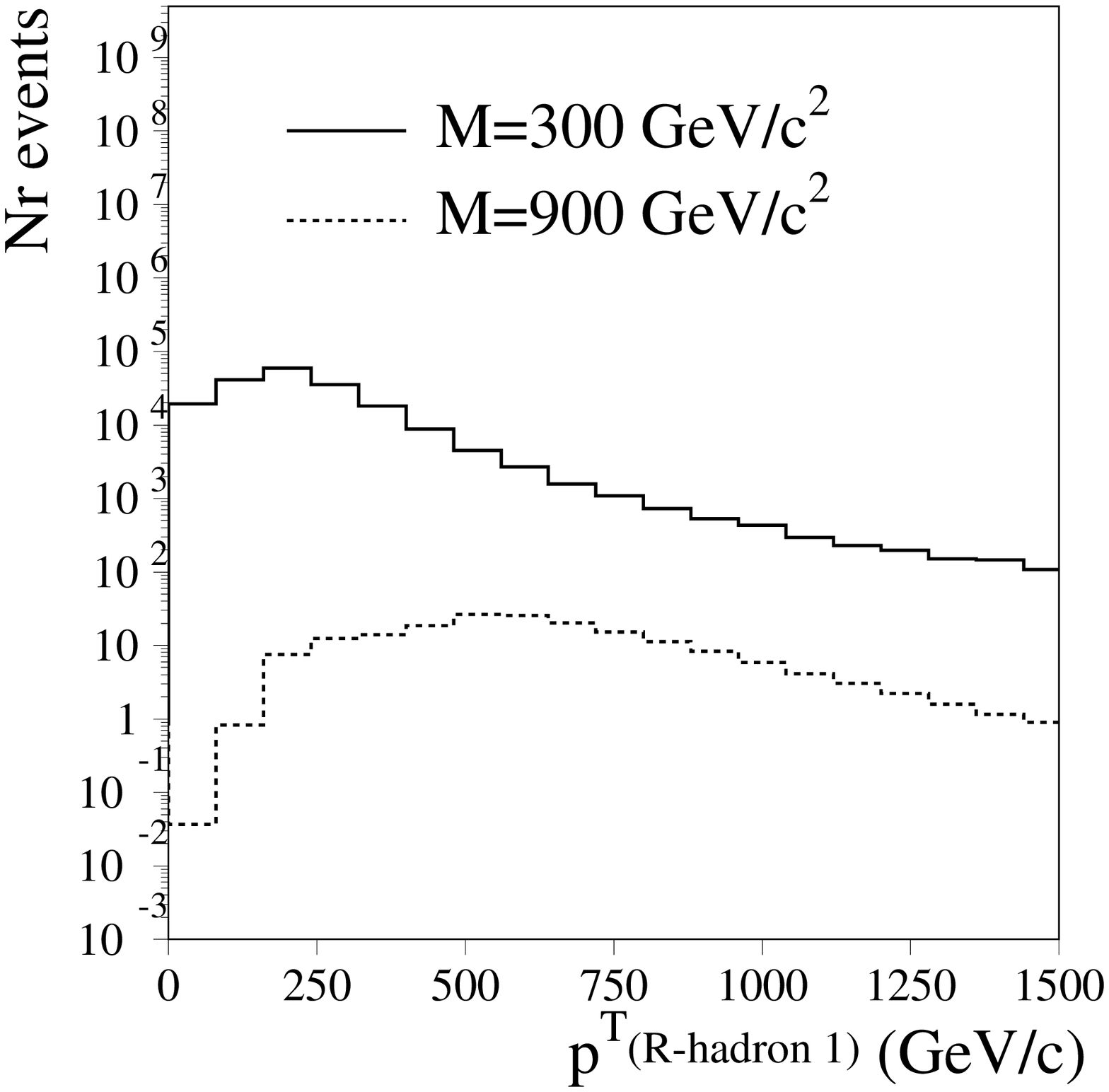,height=5.7cm,width=5.6cm}\\
\end{center}
\caption[]
{\protect \footnotesize  The missing transverse energy and the total visible energy, after high level trigger requirements for background (top) and signal (R-hadrons with masses of 300 and 900~\gevcc). The number of events corresponds to an integrated luminosity of 1~fb$^{-1}$. \label{fig:11}}

\end{figure}

\begin{figure}[b!]
\begin{center}
\epsfig{file=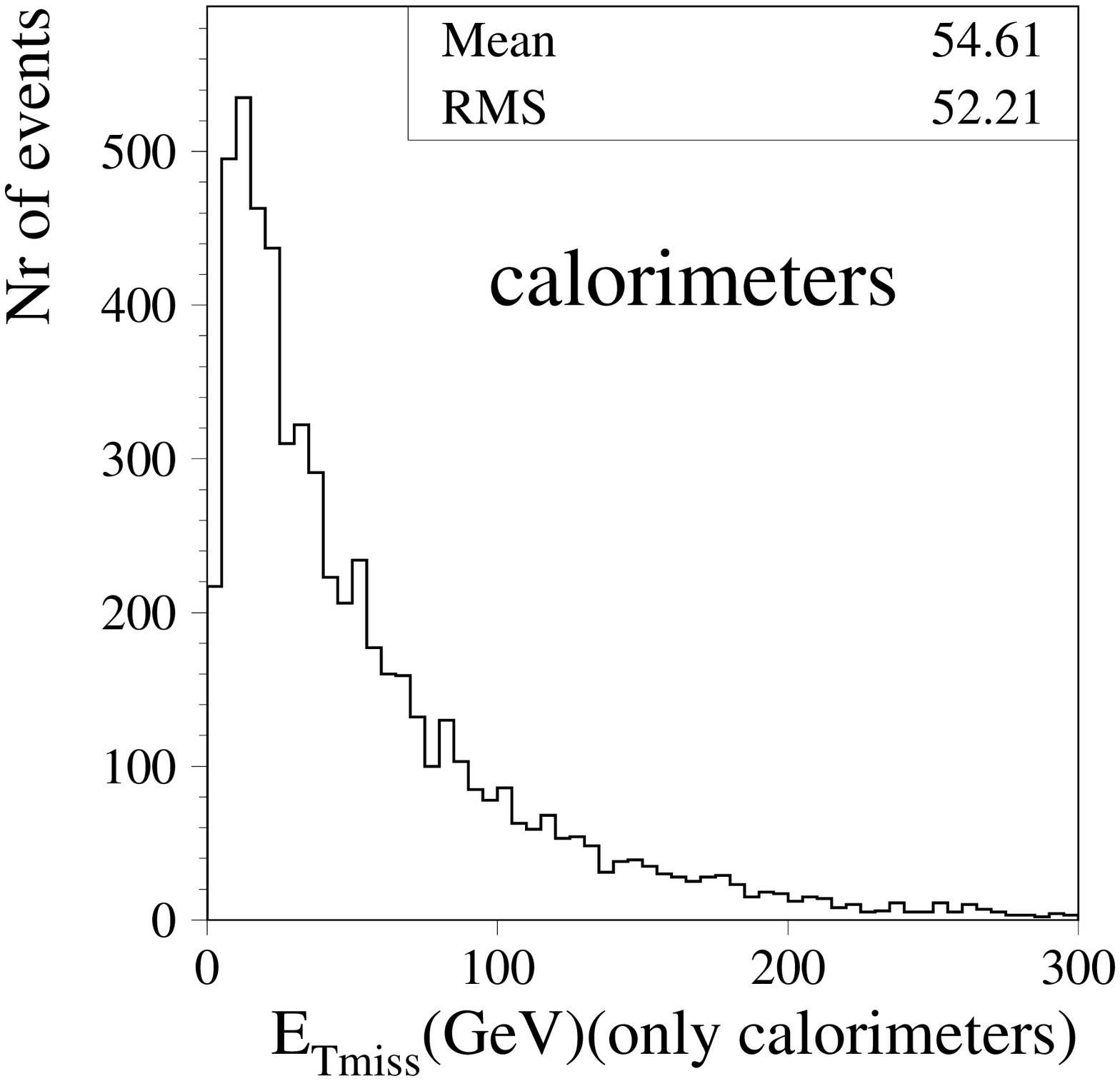,height=6.5cm,width=5.6cm}\epsfig{file=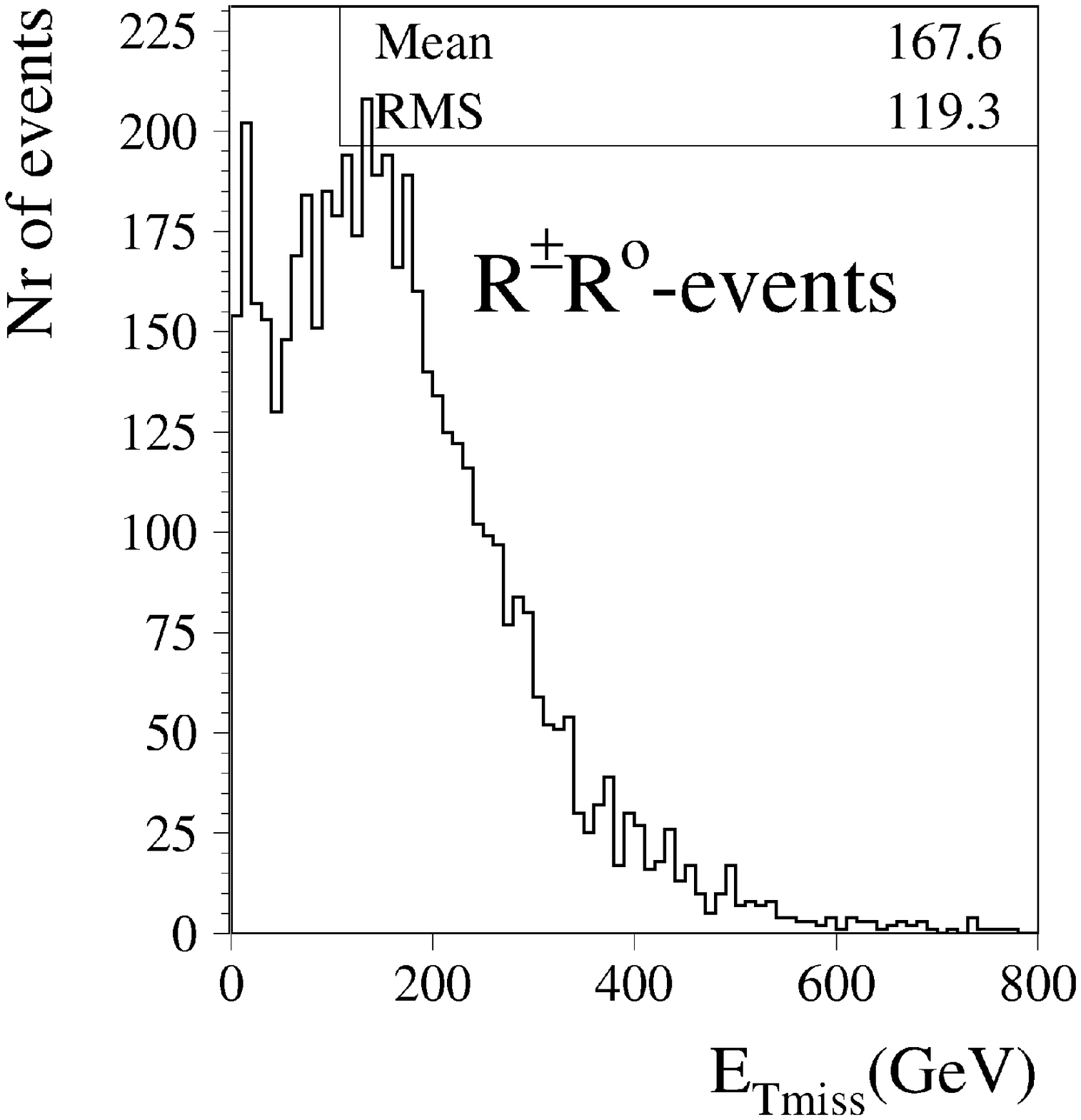,height=6.5cm,width=5.6cm}\epsfig{file=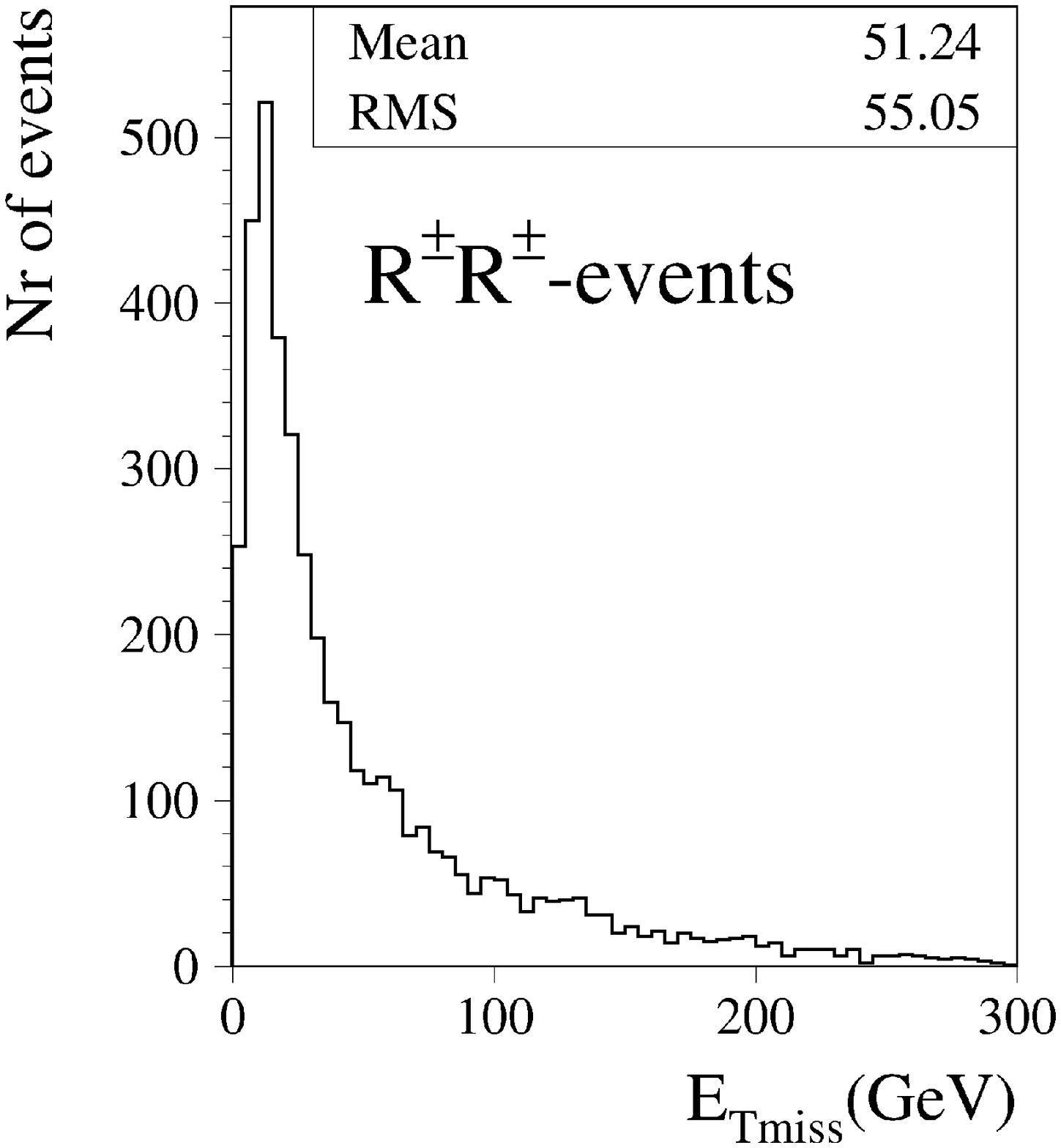,height=6.5cm,width=5.6cm}
\end{center}
\caption[]
{\protect \footnotesize  The missing transverse energy for R-hadrons
  produced via \chancc~ for R-hadron mass of 300~GeV/c$^2$ in case only
  calorimeter information is used (left), in case a neutral and a
  charged R-hadron are detected (center), and in case two charged
  R-hadrons (right). \label{fig:33}}
\end{figure}

Three variables related to the global event topology have been investigated.
\renewcommand{\descriptionlabel}[1]%
{\hspace{\labelsep}#1}
\begin{description}
\item[\em The transverse momentum in the muon chambers:]
The most energetic track reconstructed by the muon system is taken as an R-hadron candidate, and an event is required to have at least one candidate with very high transverse momentum ($P_T^{R}$).
\item[\em The missing transverse energy, $E^{miss}_T$:] A vector sum of the transverse momenta of all particles, calculated using full event information, i.e. also taking into account the inner detector and the muon chambers. For charged R-hadron candidates detected in the inner detector, the $p_T$ of the track is counted in the vector sum, while for R-hadron candidates with no track in the inner detector, the transverse energy as measured in the calorimeter is counted. However for an R-hadron candidate detected in the muon chambers but not in the inner detector, the $p_T$ as measured in the muon chambers is counted. If it is detected in both, matching is assumed possible, and only the inner detector is counted in that case.  
\item[\em The total visible transverse energy of the event, $E^{sum}_T$:] A scalar sum over transverse energies of all particles, using the same event information content as $E^{miss}_T$.
\end{description}
Distributions of the three variables for signal and background after high level trigger cuts are shown in Fig.~\ref{fig:11}. Figure~\ref{fig:33} shows the $E^{miss}_T$ distribution for R-hadron events with different reconstructed detector information. Note the significant difference in the average $E^{miss}_T$. For events where both R-hadrons have no track or muon information, only the calorimeter information is available (like the first level trigger), yielding a moderate $E^{miss}_T$. In contrast, events where the transverse momentum of only one of the two R-hadron candidates is measured by the inner detector or muon system, show a large $E^{miss}_T$, due to the inclusion of the one extra unbalanced high $p_T$ track. The case where both R-hadron candidates are fully measured, either by the inner detector or muon system, has again a moderate $E^{miss}_T$ similar to the first case.

\subsection{Time-of-flight selection variable}\label{tofvariable}
The difference in time-of-flight to the muon chambers, $\Delta$TOF, between an R-hadron candidate and a muon traveling at the speed of light, is a more involved variable, requiring a modified track reconstruction in the muon chambers. An estimate of the accuracy of the ATLAS muon chambers to measure time-of-flight as well as velocity is given in the following.

Time-of-flight measurements in ATLAS are possible due to the
staggered arrangement of the MDT's (Monitoring Drift Tubes) in the muon chambers 
and an excellent timing precision ($\simeq$1~ns) of the MDT's. 

The multiple repeated measurements of a trajectory from the staggered tube arrangement, allow a determination of the velocity of R-hadrons, by using the method of Ref.~\cite{giacomop}, which is based on $\chi^2$ minimization
w.r.t. $\beta=v/c$ of a track.
%The tube staggering the $\chi^2$ of the fitted trajectory is sensitive to the particle velocity, thus providing a tool to reconstruct $\beta=v/c$~\cite{giacomop}. Following the method of Ref.~\cite{giacomop}, based on $\chi^2$ minimization w.r.t. $\beta$ of a track, the velocity of R-hadrons is determined using the muon system.
The resulting resolution of the
R-hadron velocity in the production vertex is found to be
\begin{equation}\label{eq:sigmabeta}
\sigma(\beta)/\beta^2=0.067\pm0.029.
\end{equation}

It must be noted that R-hadron energy losses in the calorimeter 
result in a significantly smaller velocity at the muon chambers than 
that at the primary vertex. As a result of the fluctuations in energy loss, the error of the production velocity prediction is increased comparing to the velocity measured in the muon system itself.

\begin{figure}[t!]
\begin{center}\epsfig{
file=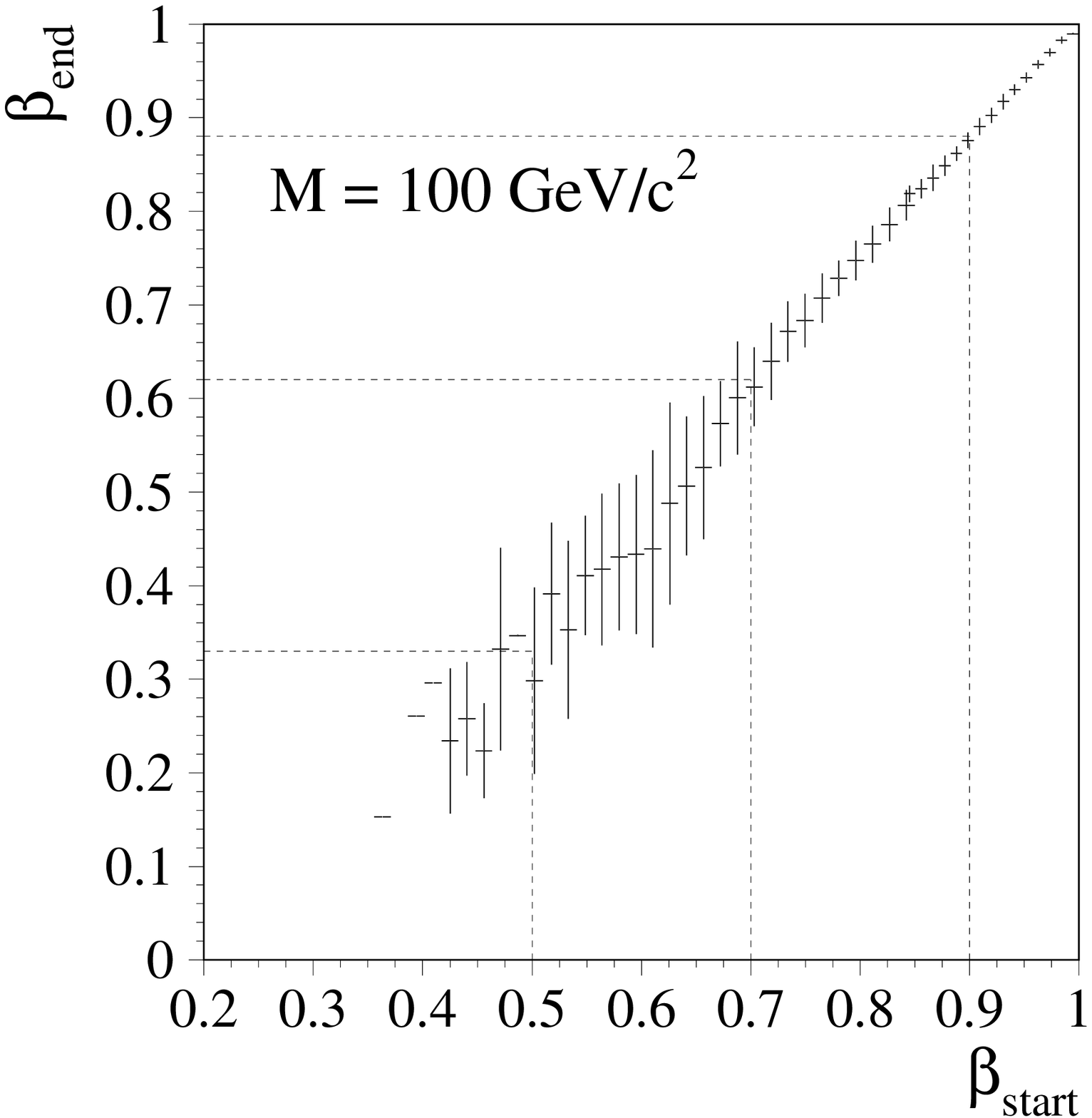,height=5.5cm,width=5.5cm}\epsfig{
file=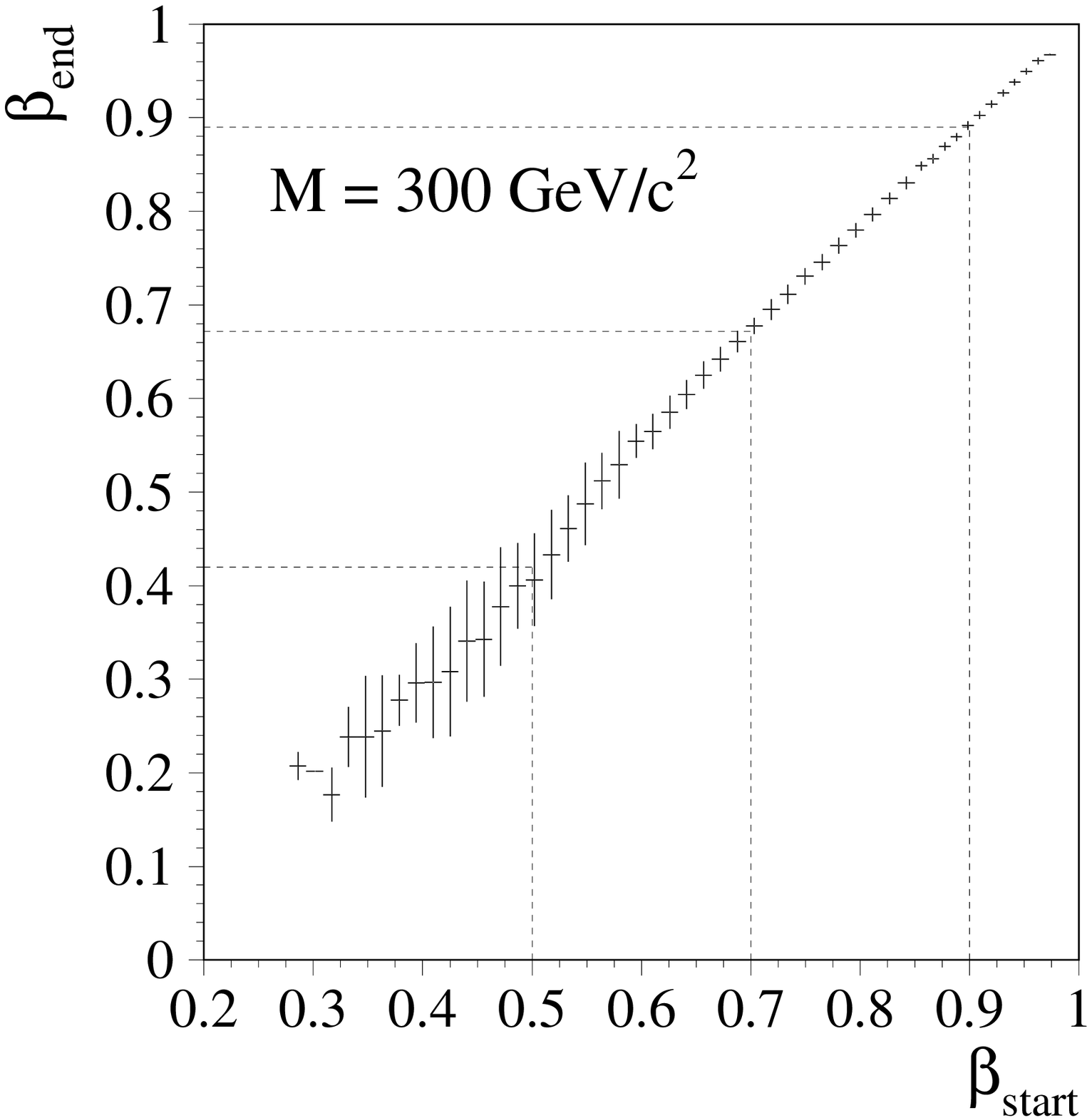,height=5.5cm,width=5.5cm}\epsfig{
file=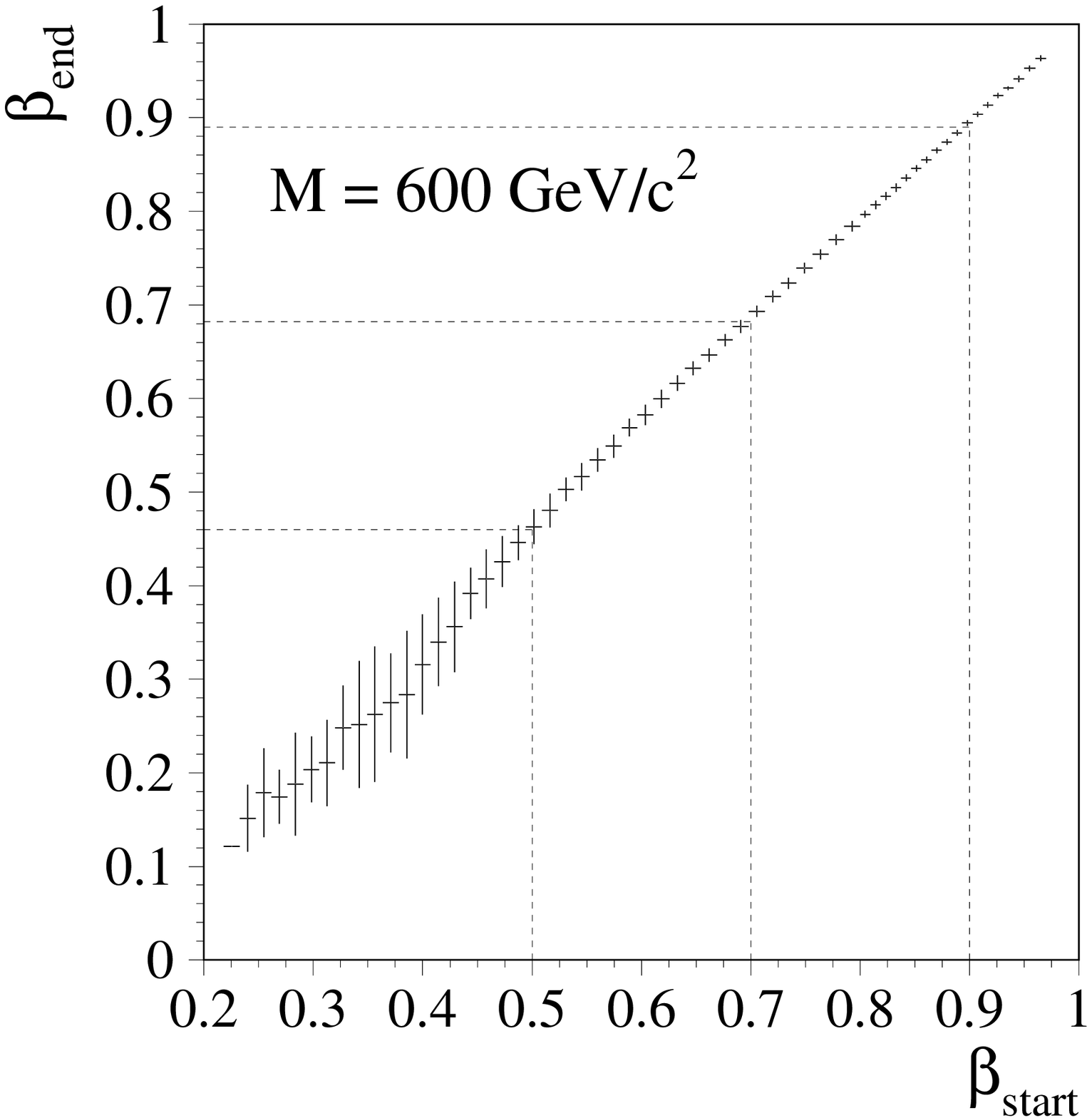,height=5.5cm,width=5.5cm}
\end{center}
\caption{\protect\footnotesize 
 The velocity $\beta_{end}$ for different R-hadron masses, as measured at the second layer 
of MDT's (7.11 m) in the central barrel region ($\eta=0.1$), as function 
of the velocity $\beta_{start}$ at the vertex. 
The error indicate the spread on the mean value.
\label{muonbe}}
\end{figure}

\begin{figure}[h!]
\begin{center}
\epsfig{file=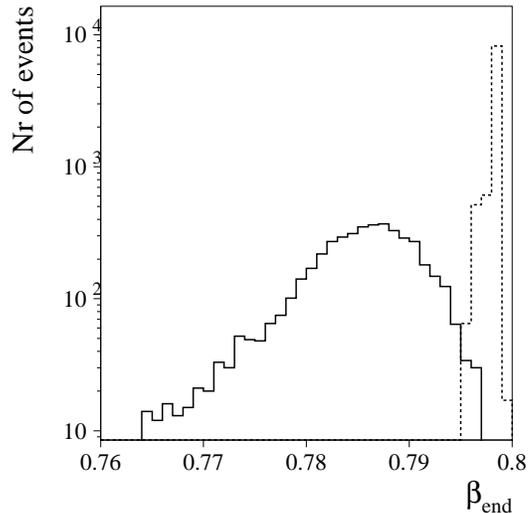,height=7.5cm,width=7.5cm}
\end{center}
\caption{\protect\footnotesize 
The distribution of $\beta_{end}$ as measured at the first layer of MDT's 
(6.92 m) when $\beta_{start}=0.8$, for a strongly interacting particle
(solid), and for a particle only interacting electromagnetically (dashed), both of mass 300~\gevcc. The spread in
$\beta_{end}$ is much larger for a strongly interacting particle, as
expected. 
\label{fig:beta1d}}
\end{figure}

In Fig.~\ref{muonbe} the velocity of  R-hadrons at the second 
(middle) layer of MDT's ($r=7.11$ m), $\beta_{end}$,
is shown as function of the $\beta_{start}$, 
the velocity at the vertex. A distribution of the
velocity at the muon chambers is shown in Fig.~\ref{fig:beta1d} for an
R-hadron and a particle without hadronic losses.
%Thus, as a result of the fluctuations in energy loss for a hadronic particle, the error of the production velocity prediction is increased comparing to the velocity measured in the muon system itself.

The resolution of the velocity measurement by the muon chambers of R-hadrons given above is used to parameterize the transverse momentum measurements by the muon chambers. In the following, we also use the $\simeq$1~ns timing precision of the ATLAS muon system directly by cutting on the measured time-of-flight of R-hadrons compared to that of muons.
 
\subsection{Selection procedure}\label{sec:atlascuts}
The definition of the signal significance, $P$, follows that of the ATLAS physics performance study~\cite{atlastdr},
\begin{equation}
P=\frac{S}{\sqrt{B}},
\end{equation}
where $S$ and $B$ are the number of signal and background events,
respectively. The usual criterion for discovery is a signal
significance greater than 5 and at least 10 observed signal events. 
Different cuts apply for different R-hadron mass samples. In Table~\ref{tab:cutssum1}, the cuts for all mass samples are summarized. The corresponding selected number of events for signal and background as well as the $S/\sqrt{B}$ ratios are listed in Table~\ref{tab:cutssum2}. 
\begin{table}[t]
\footnotesize\begin{center}
\begin{tabular}{|c||c|c|c||}
\hline
\hline
{\hbox{\lower 0.4cm \hbox{Mass}}} & 
\multicolumn{3}{|c|}{{\hbox{\lower 0.3cm \hbox{Cut}}}}\\
\cline{2-4}
\rule{0pt}{14 pt} 
      & \parbox{1cm}{\vspace{0.1cm}$E_{miss}^T$ (GeV)}& \parbox{0.9cm}{\vspace{0.1cm}$E_{sum}^T$ (GeV)} & \parbox{1.3cm}{\vspace{0.1cm}$p_T^{R}$ $(\gevc)$}  \\ \hline\hline
\rule{0pt}{14 pt} 
100   & $>40$    &  $>200$     & $>70$  \\ \hline
\rule{0pt}{14 pt} 
300  &  $>100$   &$>250$    &  $>135$  \\ \hline
\rule{0pt}{14 pt} 
600  &  $>160$  & $>450$    &  $>200$ \\ \hline
\rule{0pt}{14 pt} 
900  & $>250$  & $>600$    &  $>280$  \\ \hline
\rule{0pt}{14 pt} 
1100 & $>330$  &  $>750$   &  $>300$  \\ \hline
\rule{0pt}{14 pt} 
1300 &$>380$    &  $>900$   &  $>350$ \\ \hline
\rule{0pt}{14 pt} 
1500 & $>430$   & $>1050$ & $>430$ \\ \hline
\rule{0pt}{14 pt} 
1700 & $>470$   & $>1200$ & $>500$ \\ \hline
\rule{0pt}{14 pt} 
1900 & $>490$   & $>1350$ & $>580$ \\ \hline
\end{tabular}   
\end{center}
\caption{Selections for the different R-hadron masses.~\label{tab:cutssum1}}
\end{table}

\begin{table}[tb]
\footnotesize\begin{center}
\begin{tabular}{|c||c|c|c|c|c|c|c|c|c|}
\hline
\hline
{\hbox{\lower 0.4cm \hbox{Mass}}} & 
\multicolumn{6}{|c|}{{\hbox{\lower 0.3cm \hbox{Selected background (1 fb$^{-1}$)}}}} &
{\hbox{\lower 0.3cm \hbox{\parbox{1.2cm}{\vspace{0.2cm}Selected \hspace*{0.2cm}signal (1 fb$^{-1}$)\vspace*{-0.4cm}}}}} & \parbox{1.2cm}{\vspace{0.5cm}$S/\sqrt{B}$ (1 fb$^{-1}$)\vspace*{-0.6cm}} & \parbox{1.3cm}{\vspace{0.5cm}$S/\sqrt{B}$ (30 fb$^{-1}$)\vspace*{-0.6cm}} \\
\cline{2-7}
\rule{0pt}{14 pt} 
 $(\gevcc)$      &  QCD & b$\bar{\rm b}$ & t$\bar{\rm t}$ & W& Z &\parbox{1.15cm}{WW/ WZ/ZZ}  &   &  & \\ \hline\hline
\rule{0pt}{14 pt} 
100   & 1.3$\times 10^{4}$&1.1$\times 10^{3}$ &1.2$\times 10^{4}$ & 1.7$\times 10^{4}$&  8.2$\times 10^{2}$   & 4.5$\times 10^{2}$ & 8.7$\times 10^{6}$ & 4.1$\times 10^{4}$ & 22$\times 10^{4}$\\ \hline
\rule{0pt}{14 pt} 
300  &   6.2$\times 10^{2}$& 66   & 7.3$\times 10^{2}$  & 2.0$\times 10^{3}$ & 37 &  51 & 8.6$\times 10^{4}$ & 1.4$\times 10^{3}$ & 7.7$\times 10^{3}$\\ \hline
\rule{0pt}{14 pt} 
600  &  40&  6.8  &  85&  47  & 65 & 11  & 1.7$\times 10^{3}$ &1.2$\times 10^{2}$ &6.6$\times 10^{2}$ \\ \hline
\rule{0pt}{14 pt} 
900  & 3.1&  0.6  &  11&  13  & 2.0 & 3.2  & 97 & 17.1 & 93.7\\ \hline
\rule{0pt}{14 pt} 
1100 & 1.2  &0.2 & 4.0   &5.3  & 1.2 & 1.3 & 19  & 5.3 &29.0\\ \hline
\rule{0pt}{14 pt} 
1300 & 0.8  & 0.1 & 1.9   & 2.3  & 0.8 & 0.8 & 3.9  & 1.5 &8.21\\ \hline
\rule{0pt}{14 pt} 
1500 & 0.2  &0.1 & 0.8   &1.2  & 0.5 & 0.5 &1.0 & 0.6 &3.29\\ \hline
\rule{0pt}{14 pt} 
1700 & 0.1  & 0.0 & 0.5  & 0.7 & 0.4 & 0.2 & 0.25  & 0.18 &0.99\\ \hline
\rule{0pt}{14 pt} 
1900 & 0.1  & 0.0 & 0.5   & 0.5  & 0.3  & 0.2   & 0.05 & 0.04&0.22\\ \hline
\end{tabular}   
\end{center}
\caption{Expected number of signal and background events for the selection presented in Table~\ref{tab:cutssum1}.\label{tab:cutssum2}}
\end{table}
\begin{table}[h!]
\footnotesize\begin{center}
\begin{tabular}{|c|c|c|c|}
\hline
\hline
Mass & Selected signal (1 fb$^{-1}$) &$S/\sqrt{B}$(1 fb$^{-1}$) &$S/\sqrt{B}$(30 fb$^{-1}$)\\ \hline\hline
\rule{0pt}{14 pt} 
100   &1.4$\times 10^{7}$&  5.9$\times 10^{5}$&  32$\times 10^{5}$\\ \hline
\rule{0pt}{14 pt} 
300  &1.4$\times 10^{5}$&   1.9$\times 10^{4}$&   10$\times 10^{4}$\\ \hline
\rule{0pt}{14 pt} 
600  &2.8$\times 10^{3}$&  9.8$\times 10^{2}$&  54$\times 10^{2}$\\ \hline
\rule{0pt}{14 pt} 
900  &168&94.0& 515\\ \hline
\rule{0pt}{14 pt} 
1100 &33&42.0& 230\\ \hline
\rule{0pt}{14 pt} 
1300 &7.0&11.8&65.0\\ \hline
\rule{0pt}{14 pt} 
1500 &1.87& 4.24&23.2\\ \hline
\rule{0pt}{14 pt} 
1700 &0.44&  1.22 &6.68\\ \hline
\rule{0pt}{14 pt} 
1900 &0.08&  0.24&1.31\\ \hline
\end{tabular}   
\end{center}
\caption{Ratio $S/\sqrt{B}$ for the different R-hadron masses when
  replacing the cut on $E^T_{miss}$ and $E^T_{sum}$ by a time-of-flight cut $\Delta\rm TOF>3$~ns.\label{tab:cutssum3}}
\end{table}

These cuts are all related to the global shape and kinematics of the events, i.e. on information available at high level trigger. 

%Another selection procedure, requiring a more detailed knowledge of the
%detector and the event reconstruction as the method described above,
%would be to measure the velocity of the R-hadron candidates using the
%muon chambers, and to extract the time-of-flight information. The
%intrinsic time resolution of the muon chambers is about 1~ns~\cite{giacomop}.

Requiring the $\Delta$TOF with respect to a
particle with $\beta=1$ to be larger than 3~ns (conservatively $\gtrsim 3\sigma$)
yields a powerful rejection against muons. Thus, for this method the
$\Delta$TOF replaces the cuts on missing energy and total transverse
energy sum, while the cut on the muon transverse momentum is
unchanged. For heavy R-hadrons (masses above 1~\tevcc) the $\beta$ is
typically 0.8 or smaller and a tighter cut on the time delay could be used, but the limiting factor
for discovery is the gluino production cross-section. Table~\ref{tab:cutssum3} shows the corresponding
signal significances.

\section{Results and discussion}~\label{sec:atlasconc}
Using only global event variables, available from standard high level
event reconstruction, Table~\ref{tab:cutssum2} shows that R-hadrons can be discovered for masses up to
1400~\gevcc\ for an integrated luminosity of 30~fb$^{-1}$ assuming low luminosity muon trigger configurations. In
particular, for low masses, a signal significance of 5, the usual
criterion for discovery, could be reached after only a few days of
running. Including more specialized information from a dedicated
reconstruction, such as time-of-flight, improves the sensitivity and
pushes the discovery limit for R-hadron masses up to 1700~\gevcc. The
numbers in Table~\ref{tab:cutssum2} and \ref{tab:cutssum3} represent conservative estimates,
as we do not include the model-dependent contributions from $\rm
q\bar{\rm q}\rightarrow \tilde{\rm g}\tilde{\rm g}$, which, for high
gluino masses, can be larger than 100\%, see
Fig.~\ref{fig:rhadprod}a. For example, in the framework of split
supersymmetry with very high squark masses, the discovery potential
would extend up to masses of 1600~\gevcc\ if only using global event
shape variables, and up to 1800~\gevcc\ if time-of-flight information would be included.

Although the high luminosity muon trigger configuration has not be studied, the reduction in muon trigger efficiency is expected to be small, as seen in Table~\ref{tab:efftrig}. A reduction of less than 50~\gevcc\ in discovery potential is expected.

There are a number of uncertainties that may influence the precise
discovery limit for R-hadrons and in the following we discuss the most
important sources.

The probability to obtain a gluino-gluon state in the hadronization
process, $P_{\tilde{\rm g} \rm g}$, by default taken to be 0.1,
is directly related to the ratio of charged to neutral R-mesons produced in the hadronization process. The extreme value of 1.0 implies that only
neutral R-hadrons are produced. Due to the repeated charge flipping of the
R-hadrons as they traverse the calorimeters, the R-hadron signature
in the muon system is independent of the initial charge, hence, a
selection based on muon-like signatures is unaffected by this
parameter. For our selection using missing and visible energy, and thus relying on
inner detector information, the significance is at most reduced by a factor 2
for events generated with $P_{\tilde{\rm g} \rm g}=1$, as the momentum imbalance in the tracking system, as well as the total amount of visible energy, are smaller on average. A significance loss by a factor 2 lowers the discovery reach by about 100~\gevcc. Although no track is
present in the inner detector, the selection is still effective, as
the R-hadron can still be detected in the muon system and thereby
contribute to a total energy signal. Another issue is the handling of
events with neutral R-hadrons in the inner detector and charged
R-hadrons in the muon system by the event reconstruction. This
uncertainty has not been quantified, because a detailed knowledge of
the final reconstruction software is needed. But it is not expected to
be a limiting factor for discovering R-hadrons, as the software can be readapted if necessary.

As shown in Table~\ref{tab:intmuon}, R-hadrons have a non-zero
probability to undergo a nuclear interaction in the muon
system, and thereby possibly changing their charge. This probability can be as
high as roughly 50\% over the full muon system due to the presence of
support structures and cryostats. As can be seen from
Table~\ref{tab:intmuon}, a large fraction of R-hadron events would
actually fire the muon trigger, and the uncertainty arises thus mainly
from the full track reconstruction in the muon chambers of such events. The actual loss due to this effect depends on the available charge exchange processes for a given initial R-hadron charge and the impact of the lost energy in the interaction on the subsequent track reconstruction and acceptance. The worse-case scenario is that any interaction inside the full muon system implies that the R-hadrons is reconstructed badly and completely lost. In that case, the fractional losses of events containing two R-hadrons, due to one or more interactions inside the full muon system (3rd column of Table~\ref{tab:intmuon}), are approximately 38\%, 34\%\ and 31\%, for gluino masses of 100~GeV/$c^2$, 300~GeV/$c^2$ and 600~GeV/$c^2$, respectively. For a reduction of 35\% on the number of signal events, the discovery reach decreases with about 70~\gevcc.

Additional uncertainties related to the generation and simulation of
events, such as parton distribution functions, higher order
corrections and background rate uncertainties have not been
studied. Furthermore, the calibration and final understanding of the
detector response may influence the above results. No attempt has been made to quantify these
kind of uncertainties, but given the many and very distinct signatures
of R-hadrons, they are expected not to have a dramatic impact on the
abilities for discovering R-hadrons over a broad mass range. 

\section{R-hadron identification from additional detector signatures}
In the situation that an excess of events with slow, yet high $p_T$, tracks in the muon system and/or high $E_T^{miss}$, is observed, it is necessary to study additional signatures to confirm the presence of R-hadrons. In addition, the identification of R-hadrons can provide valuable information about the structure and the parameters of the underlying theory. The production rate of neutral R-hadrons and the fraction of charge flipping can determine whether the heavy spectator is a gluino or a squark and measure the probability to form \glu g bound states. In this section we present possibilities to identify and classify R-hadrons by using more detailed detector information such as the TRT, the E/p ratio, and shower profiles in the calorimeters.

These signatures may also help to improve the event selection, but the very high selection efficiency and background rejection of a time-of-flight based selection, makes it difficult to improve the selection for events with at least one well-reconstructed R-hadron candidate in the muon detector. Of the remaining events without a muon trigger, up to about 57\% (Table~\ref{tab:efftrig}) at low luminosity for high R-hadron masses, at most 50\% will pass the jet/$E_T^{miss}$ triggers (Table~\ref{tab:efftrigcal}), prior to any further event-topology or R-hadron specific selection. As seen in Section~\ref{sec:atlasconc}, a factor 2 (slightly less than for the case of split supersymmetry) increase in the number of signal events changes the discovery reach with at most 100~\gevcc.

\subsection{Signatures in the TRT detector}\label{sec:innerdet}
For events with at least one charged R-hadron produced in the hard process, the typically slow charged R-hadron will deposit large ionization energy signals
in the tracking system. Since the TRT has been designed for dE/dx measurements, this subdetector has been investigated in more detail. A similar possibility exists for the ATLAS pixel detector~\cite{idtdr}.

%\subsubsection{Signatures in the TRT detector}
%\label{sec:trtsig}
The ionization energy deposits of a slow charged particle grow approximately like $1/\beta^2$.
Hits in the ATLAS TRT detector are registered using two
discriminator levels: a low threshold level, corresponding
approximately to 200~eV, and a high threshold level, corresponding to a 
transition radiation photon of about 5~KeV. The time of the low
threshold crossing is measured with a 3.125~ns binning. With these
measurements, R-hadron identification can be achieved in two ways: by counting the number of straw hits exceeding the
high threshold (HT) discriminator level~\cite{atlastdr} or by using
the Time-over-Threshold method (ToT)~\cite{atlastdr,idtdr,chafik}.

The number of high threshold hits of R-hadrons, compared to those of electrons and muons, has been studied using fully simulated electrons, muons and R-hadrons of masses 
(100~\gevcc, 300~\gevcc, 600~\gevcc), generated at various transverse momenta and $\eta$-values.

\begin{figure}[]
\begin{picture}(0,74)
\put(15,0){\epsfxsize70mm\epsfbox{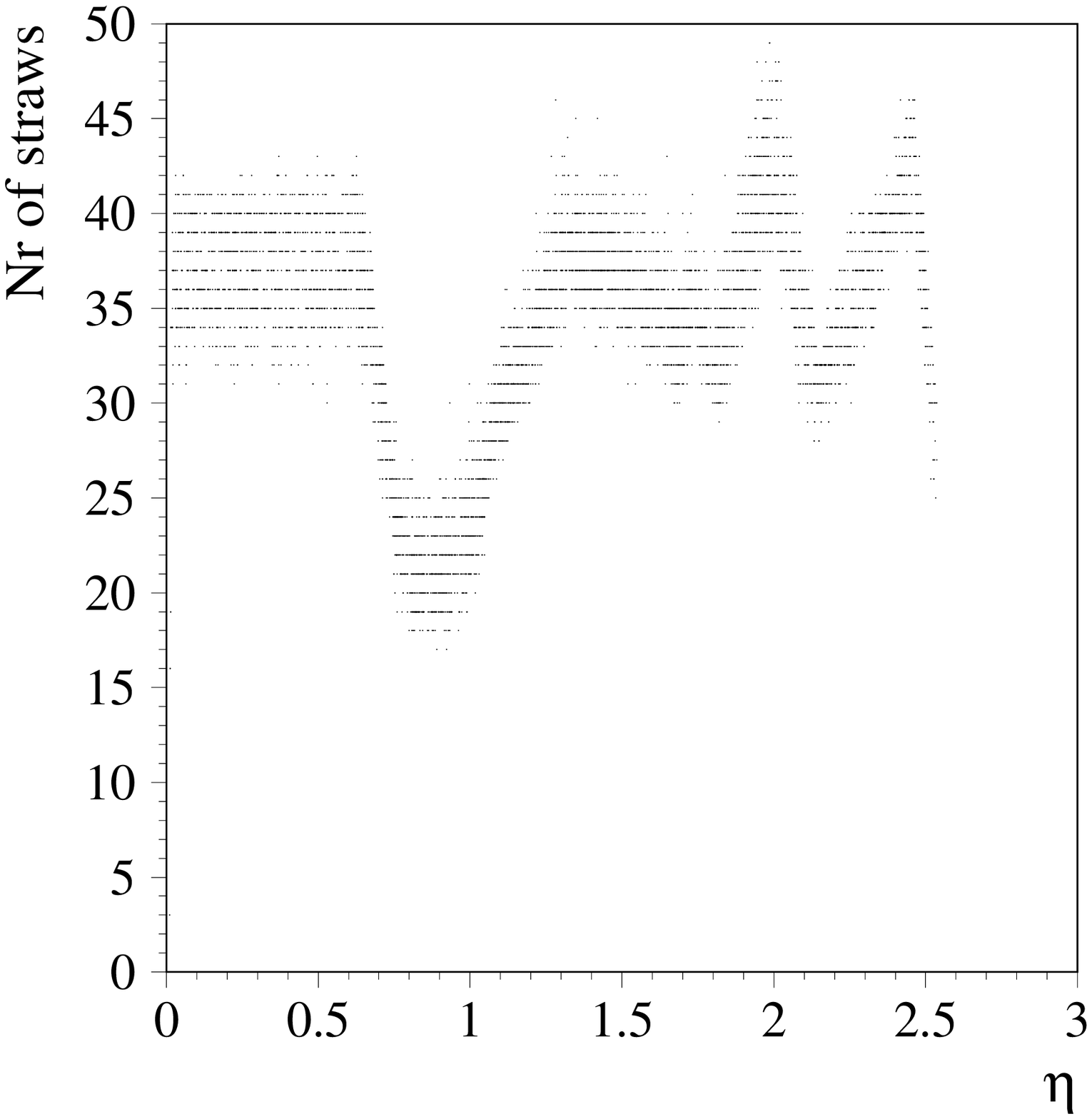}}
\put(48,31){(a)}
\put(85,0){\epsfxsize70mm\epsfbox{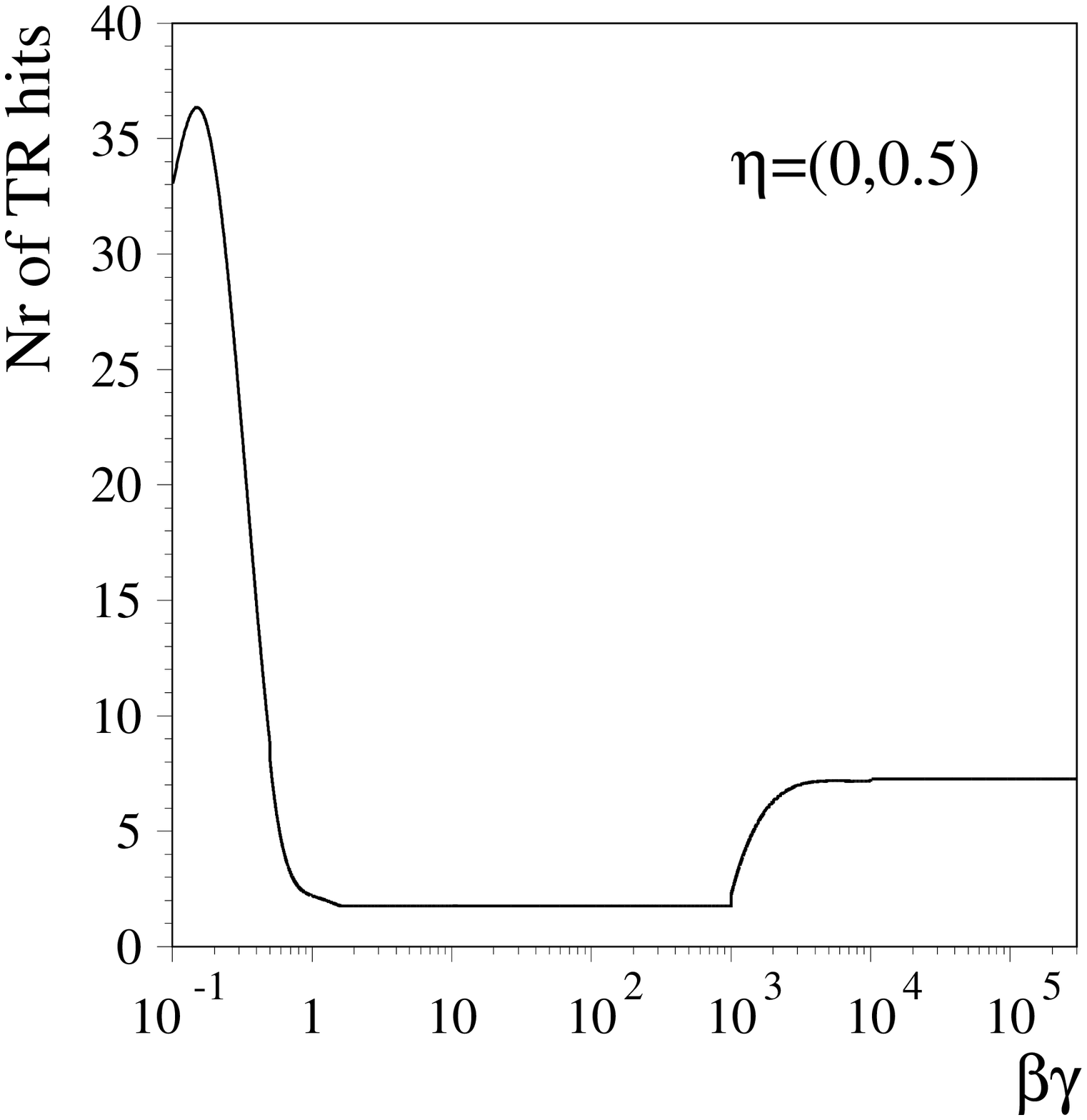}}
\put(118,31){(b)}
\end{picture}
\caption[ ]
{\protect\footnotesize
(a) The number of crossed straws as function of R-hadron pseudo-rapidity. 
(b) Average number of HT hits as function of $\beta\gamma$ of a particle 
   at central pseudo-rapidity.
\label{fig:straws}}
\end{figure}

In Fig.~\ref{fig:straws}a the number of straws crossed by a particle 
is displayed as function of $\eta$. The number of crossed straws
varies with $\eta$, and so does the the average amount of high
threshold hits. The TRT detector is therefore divided into five $\eta$
regions, and the average amount of high threshold hits is studied as
function of $\beta\gamma$ of the particle. 

\begin{figure}[h!]
\begin{picture}(0,74)
\put(15,0){\epsfxsize58mm\epsfbox{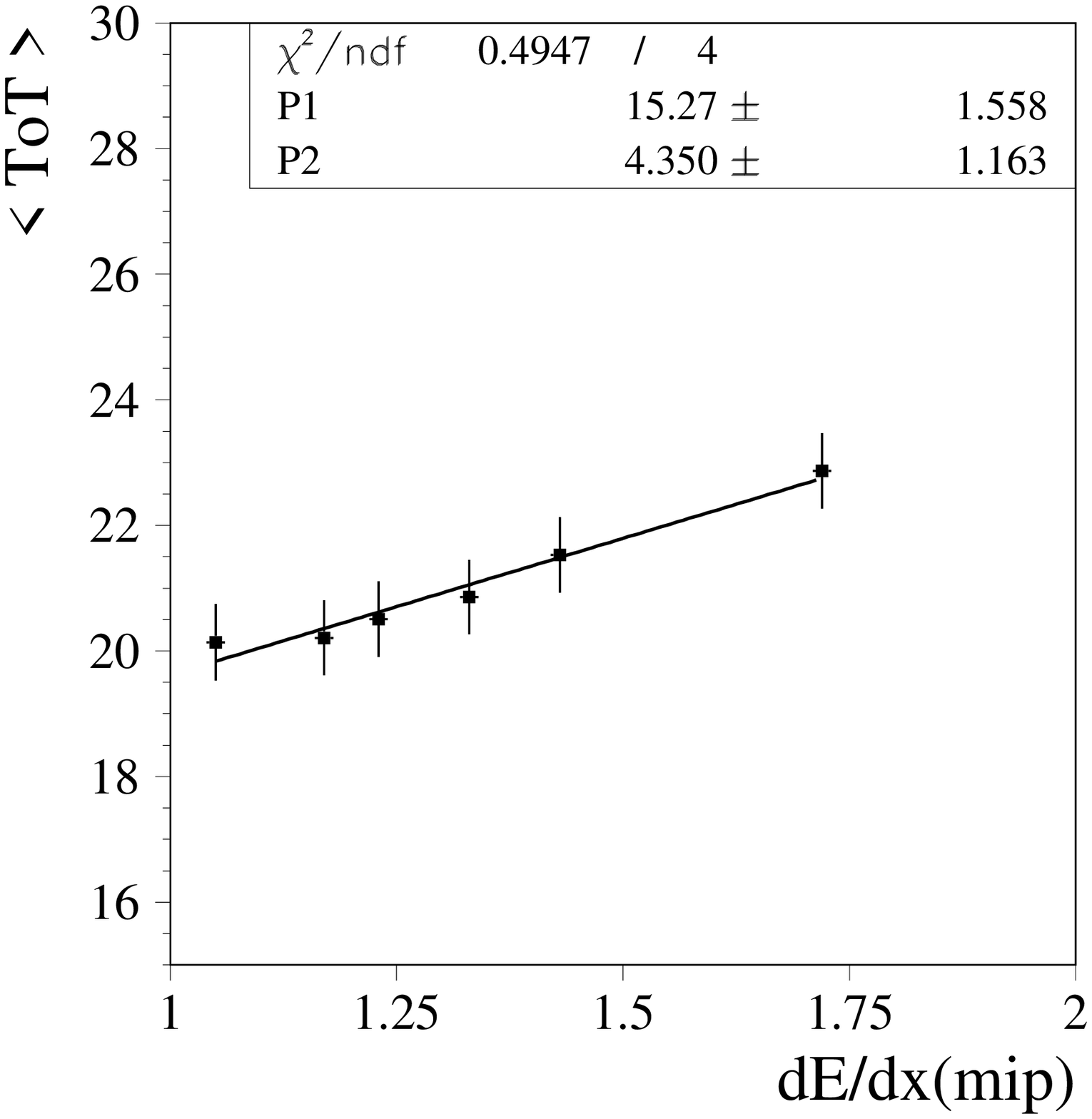}}
\put(35,31){(a)}
\put(85,0){\epsfxsize58mm\epsfbox{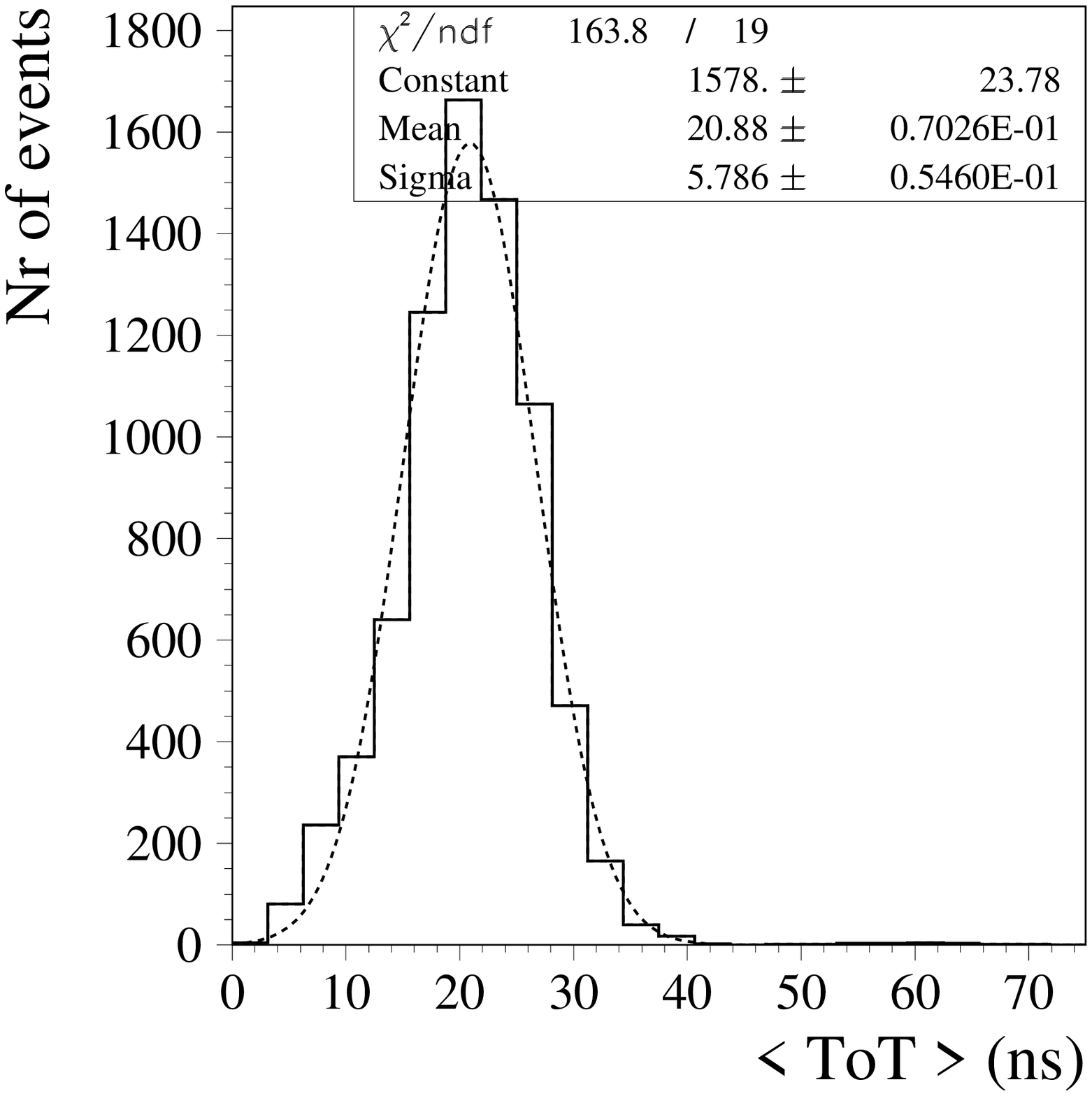}}
\put(118,31){(b)}
\end{picture}
\caption[ ]
{\protect\footnotesize
(a) The relation between the d$E$/d$x$ of a pion, and its average ToT. (b) A 1-dimensional projection of the ToT obtained from testbeam pion data (5~GeV/c)~\cite{chafik} for central values (between 0.8 and 1.2 mm) of the distance of closest approach, fitted with a Gaussian. The plot is independent of particle type and depends only on the $\beta$ of the particle. 
\label{fig:totapp}}
\end{figure} 

As it can be seen from Fig.~\ref{fig:straws}b,
HT hits can play a role for R-hadron identification 
only for $\beta\gamma\lesssim 1$, or $\beta\lesssim0.7$. 
On the other hand, the number of HT hits can be used to separate 
fast R-hadrons with $\beta\gtrsim 0.7$, when they are minimum ionizing, 
from very  high $p_T$ muons ($p_T\gtrsim200$~GeV/c) 
which start to emit transition radiation photons.

The Time-over-Threshold method (ToT)~\cite{atlastdr,idtdr,chafik} is based on measuring the time in which a signal is above the low threshold discriminator level. In general, the more energy a particle deposits, 
the longer the signal will remain above the threshold, and thus, by measuring the time elapsed between a signal exceeding and falling below the threshold, it is possible to extract information about the d$E/$d$x$: the more ionization energy a particle deposits, the longer the ToT. The TRT offers the possibility to measure the ToT\footnote{This analysis has been done using test beam data with the TRT electronics developed in 1999. Recently, the TRT electronics was changed to provide a more stable drift time measurements independent on the signal amplitude. This makes ToT measurement less sensitive to the particle d$E$/d$x$.}.
The ToT depends on the hit distance to the anode in the straw (the closer to the anode wire the track is located the larger the ToT) and on the deposited energy. The ToT averaged over all hit distances 
in a straw, denoted $<$ToT$>$, is a function of the d$E$/d$x$ following the Bethe-Bloch 
description. In Fig.~\ref{fig:totapp}a the $<$ToT$>$ is displayed as
function of the theoretical d$E$/d$x$. 
The linear dependence between the d$E$/d$x$ and $<$ToT$>$ allows the expected $<$ToT$>$ of an R-hadron to be obtained assuming that the d$E$/d$x$ for an R-hadron follows Bethe-Bloch dependence, that is, $(\textrm{d}E/\textrm{d}x)_R \propto (\textrm{d}E/\textrm{d}x)_{(\textrm{mip})}\times (1/\beta^2)$. As is shown in Fig.~\ref{fig:totapp}b, the 
$<$ToT$>$ has a Gaussian distribution. This plot is obtained from testbeam pion data (5~GeV/c)~\cite{chafik}, but is independent of particle type and depends only on the $\beta$ of the particle.

\begin{figure}[t]
\begin{center}
\vspace{-0.7cm}
\epsfig{file=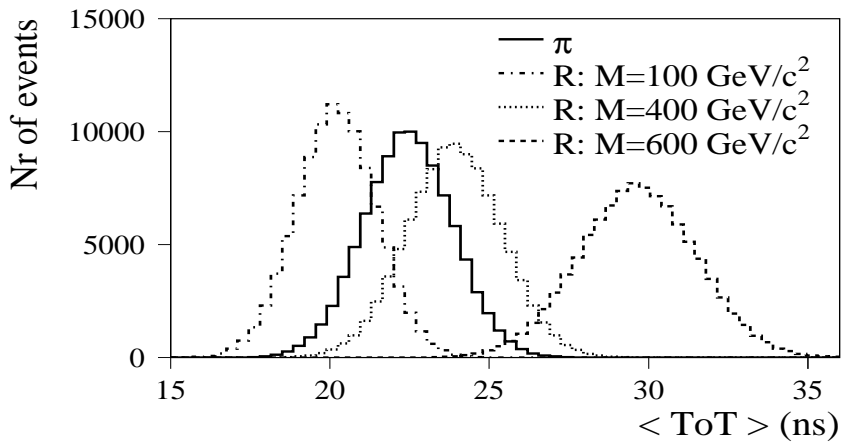,height=7.3cm,width=13cm}
\caption[ ]
{\protect\footnotesize
The average ToT for a particle hitting 33 straws (the average over all eta values) for pions of momentum 400~GeV/c (solid curve) and
for R-hadrons with momentum 400~GeV/c and mass 100~GeV/c$^2$ (or $\beta=0.95$, dashed-dotted), mass 400~GeV/c$^2$ (or $\beta=0.71$, dotted), and mass 600~GeV/c$^2$ (or $\beta=0.55$, dashed). Note that results only depend on the velocity.
 \label{fig:separation}}
\end{center}
\end{figure}

To estimate the separation power of the ToT method between heavy hadrons
and for example pions, 
a simulation experiment is done with 100000 measurements, 
for different $\beta$ values, an average ToT is generated 
and compared to that of ultrarelativistic pions.
The results depend only on the velocity and are displayed for pions and R-hadrons with momenta of 400~\gevc\ in Fig.~\ref{fig:separation}.
It is seen that for R-hadrons with velocities $\beta=0.55$, $\beta=0.71$
and $\beta=0.95$ a separation of 5.2$\sigma$, 1.0$\sigma$ and 1.6$\sigma$, respectively, 
can be obtained. Note that for high $\beta$ values the R-hadron becomes less ionizing 
than an ultra-relativistic pion.

The number of high threshold hits and the $<$ToT$>$ in the TRT are 
the only estimators of d$E$/d$x$ available in ATLAS. 
These estimators are very highly correlated, so that the separation 
power of the two together is only marginally larger than the separation 
power of the best single estimator. 

\subsection{Matching inner detector and calorimeters}
The low energy deposits from an R-hadron passing the calorimeters (Fig.~\ref{fig:en}) makes the comparison of the momentum measurement of the track with energy deposits a potential useful separation variable against pions and electrons. A simulation has been made to match inner detector momenta with calorimeter deposits of fully GEANT 3 simulated single particles.  
\begin{figure}[b!]
\begin{center}
\epsfig{file=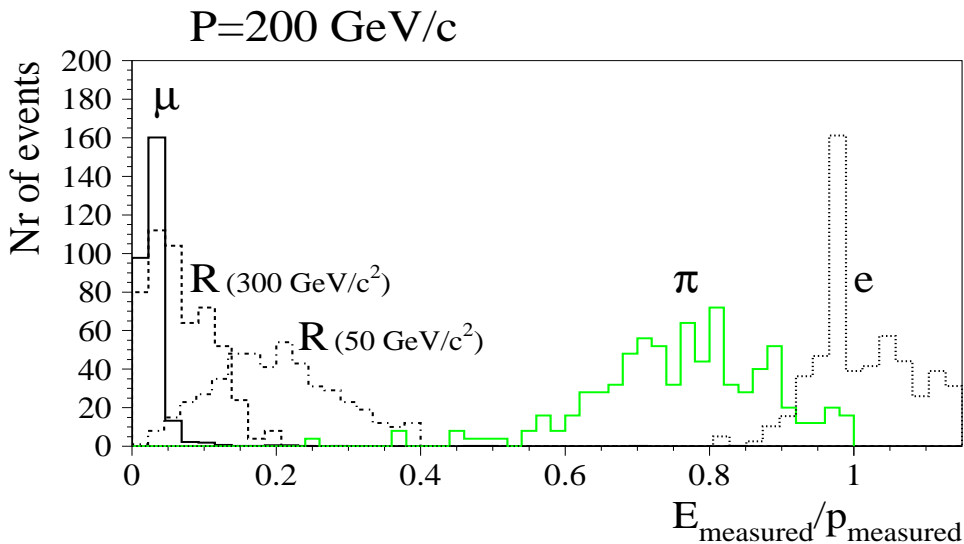,height=7.4cm,width=14cm}
\caption[ ]
{\protect\footnotesize  The ratio $E/p$ for R-hadrons, muons, pions and electrons in the barrel region ($\eta=0.1$). Singly charged R-mesons, muons, pions and electrons are generated and reconstructed. The two extreme cases are represented by the muons (solid line) and the electrons (dotted line).
\label{fig:ep}}
\end{center}
\end{figure}
In Fig.~\ref{fig:ep}, the ratio of the measured energy in the
calorimeters and the momentum of the track, $E/p$, is displayed for single R-hadrons, muons, electrons and pions in the barrel region. The performance of this method improves for high R-hadron masses due to the lower average $\beta$. A quantitative study with fully simulated R-hadron events is required to understand the interplay from nearby particles in the subsequent reconstruction. Such a simulation is not yet available.

\subsection{Signatures in the calorimeters}\label{sigcalo}
Due to the charge flipping nature of R-hadron nuclear interactions and the model dependent production rate of neutral R-hadrons, the only \emph{certain} signature of an R-hadron is that it deposits a somewhat moderate amount of energy in the ATLAS calorimeters. Therefore, calorimeter measurements could be used in events containing only neutral R-hadrons, or to confirm that the particle detected in the muon chambers is a hadron, and not a muon. In addition, the calorimeters can also be used to provide time-of-flight information like the muon system. However, given the low energy deposits and the delayed arrival of slow R-Hadrons, the time determination will have a larger uncertainty. An additional uncertainty is the propagation time on the shower development for an R-hadron, which has not been available in the GEANT3 study.

Although the precise hadronic energy deposit is subject to many uncertainties~\cite{aafke}, a few general observations can be made concerning the shower profiles.

\subsubsection{Longitudinal calorimeter profile}
The longitudinal energy deposit profile for fully GEANT 3 simulated R-hadrons has been studied and compared to those of electrons and pions in the different calorimeter compartments of the ALTAS detector.

Differences are due to the fact that the typical energy loss per interaction for R-hadrons is small relative to the R-hadrons energy.  
Contrary to energy losses of electrons or pions, the R-hadron energy deposits remain fairly constant along the trajectory of the particle, however it should be emphasized that this is only approximately true when nuclear interactions are dominating; for low $\beta$ (e.g.\ when particle is significantly slowed down) the usual electromagnetic energy loss is dominating. Since R-hadrons typically are produced with high momenta relative to their mass and thus high kinetic energy, the constant energy deposit approximation generally holds, and the energy deposit is linearly proportional to the thickness of a calorimeter compartment. In this aspect R-hadron profiles are similar to those of minimum ionizing particles like muons, apart from the absolute scale of the energy deposits. For example, an R-hadron of mass 300~\gevcc\ and momentum 200~\gevc, the averaged energy deposits in the tile calorimeter compartments amount to 2.5, 6.2 and 3.0~\gevcc, for corresponding thicknesses of 1.4, 3.9 and 1.8 interaction lengths, respectively. Note, however, that although the \emph{average} profiles for muons and R-hadrons are similar, the event-by-event variation for R-hadrons, due to charge-flipping and fluctuations in the energy loss, are considerably larger than for muons.

\subsubsection{Lateral calorimeter profile}
The lateral energy profile is investigated and compared to that of pions, muons and electrons. The width of an R-hadron shower is determined as the fractional energy measured inside a wedge as function of its opening angle. The energy is calculated as the sum of the calibrated cell energies in the ATLAS calorimeters.
\begin{figure}[t!]
\begin{center}
\vspace{-1cm}
\epsfig{file=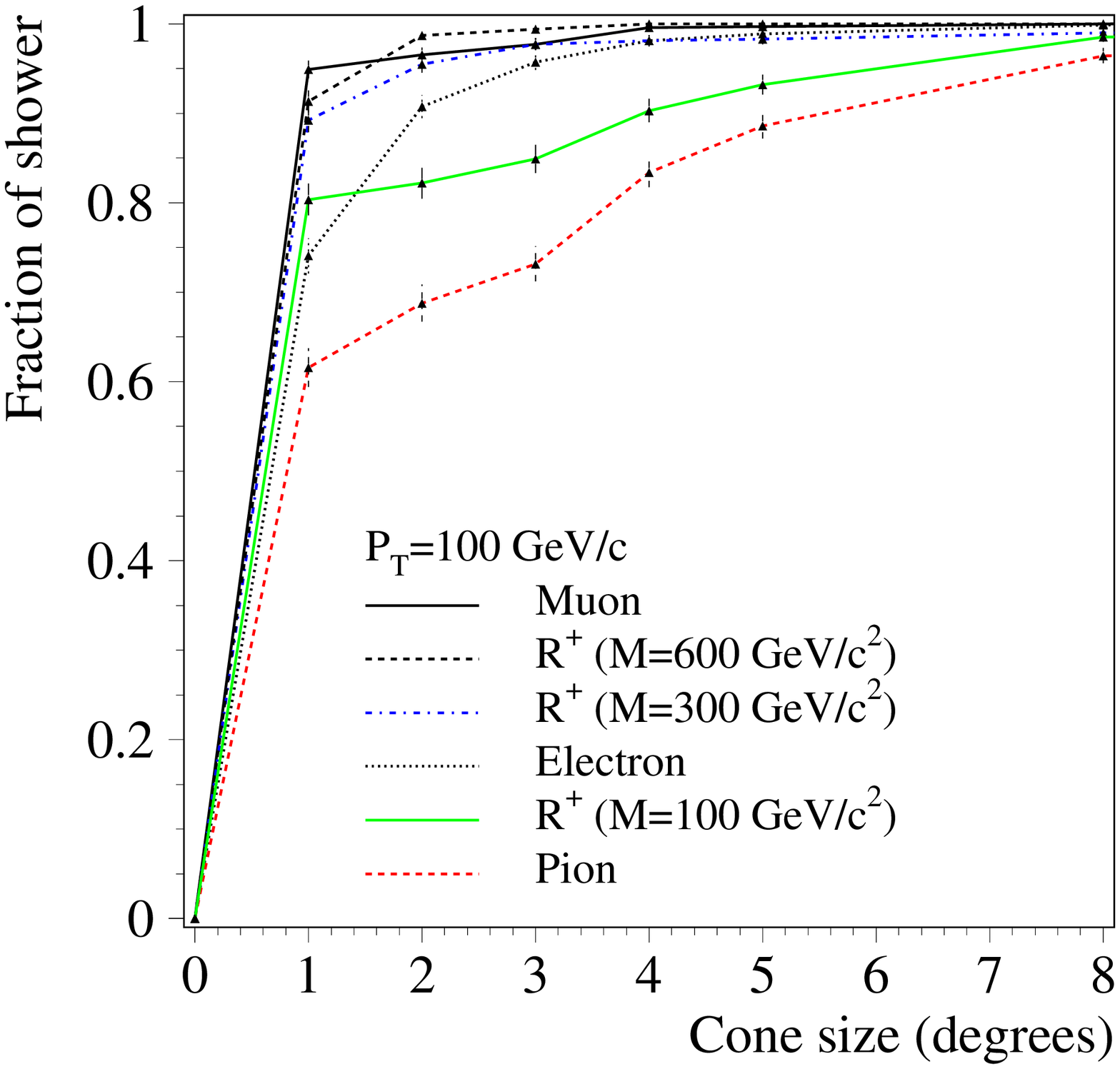,width=8cm}\epsfig{file=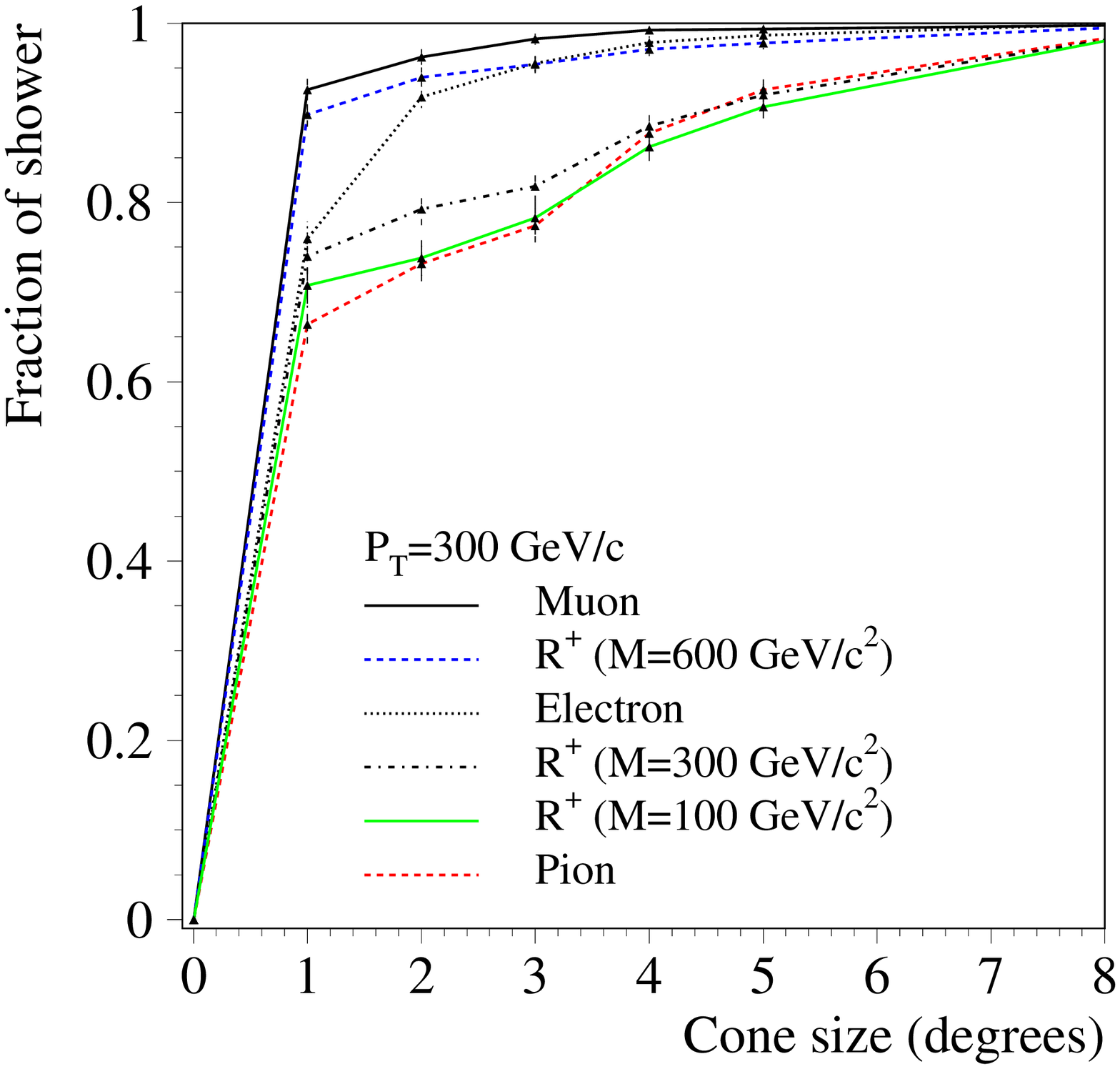,width=8cm}
\caption[ ] 
{\protect\footnotesize Lateral energy profile for muons, electrons, pions and R-hadrons of different masses with $p_T=100$ and 300~\gevc as obtained with GEANT 3. \label{fig:lateral}}
\end{center}
\end{figure}

The results are shown in Fig.~\ref{fig:lateral} for particles generated in the central detector region, with an $\eta$ value in between 0 and 0.1, and an arbitrary $\phi$ direction. Due to the nuclear interaction nature of the R-hadron shower, the shower can be quite broad, depending on the R-hadron kinetic energy. It can be seen that for R-hadrons with large kinetic energy, the shower can be as wide as that of a pion (for very high energies even broader), while slow R-hadrons exhibit showers which are as narrow as muon showers. However, the full shower shape depends highly on the amount of other particles produced in the hadronization process. Although the broadness of the shower profile could possibly be used in discriminating R-hadrons from other particles, in particular from high $p_T$ muons, a detailed parametrization and understanding of the event-by-event fluctuations of the shower width is left for a future study.

\section{Conclusion}
In this paper we have addressed many aspects important for the detection of
heavy hadronically interacting particles with exotic and complex
signatures in the detector:  R-hadrons. Using only a few general
observables and making use of standard detector techniques with the ATLAS detector would allow the discovery of
R-hadrons 
with a mass reach of at least 1400~\gevcc\ for an integrated luminosity of
30~fb$^{-1}$ at low luminosity running. R-hadrons with masses as predicted by standard SUSY scenarios (M$<$600~\gevcc) could already be discovered at very early stages of the running of the LHC accelerator.

A stand-alone precision muon detector with excellent time-of-flight
capabilities, such as the ATLAS muon spectrometer, improves
significantly the discovery reach and allows a more model-independent
search. Using only the ATLAS muon spectrometer, this paper has shown
that R-hadrons produced via \chancc\, with masses below 1700~\gevcc,
can be discovered at the LHC for an integrated luminosity of 30~fb$^{-1}$
at low luminosity running. Including R-hadrons produced via \chandd, the discovery reach of heavy gluinos, as predicted by for example split supersymmetry models, extends up to at least 1.8~\tevcc. 

At high luminosity the acceptance in the velocity,
$\beta$, of the R-hadrons is reduced, due to changes in the ATLAS muon
trigger configuration, but as the analysis can be performed
essentially on high-$p_T$ muon like signatures, only a small reduction of the discovery potential is expected.

This paper also discusses additional R-hadron detector signatures which potentially can be used to improve the discovery reach even further.

We conclude that the ATLAS detector has excellent chances of finding R-hadrons with masses as predicted by standard supersymmetry scenarios as well as those predicted by split supersymmetry models. 

\section{Acknowledgments}
This work has been performed within the ATLAS Collaboration, and we
thank collaboration members for helpful discussions. We have made use
of the physics analysis framework and tools which are the result of
collaboration-wide efforts.

The authors would like to thank Torbj\"orn Sj\"ostrand for providing
gluino hadronization routines and for many useful discussions. We
are also grateful to Aleandro Nisati for discussions about the ATLAS
muon trigger.

\end{document}